\documentclass[useAMS,usenatbib,fleqn]{mn2e}

\usepackage{graphicx}
\usepackage{float}
\usepackage{epstopdf}
\usepackage{amsmath}
\usepackage{amssymb}
\usepackage[labelformat=empty,labelsep=none]{subcaption}
\usepackage{epsfig}
\usepackage{times}

\def\gtsim {\gtrsim}   
\def\ltsim {\lesssim}   

\newcommand{\dummytitle}[1]{}

\newcommand{\msun}{{\rm M}_\odot}

\defcitealias{pt14}{TK14}
\defcitealias{pt15a}{TK15a}
\defcitealias{pt15b}{TK15b}

\title[Evolution of Scaling Relations]{Time Evolution of Galaxy Scaling Relations in Cosmological Simulations}
\author[P.~Taylor and C.~Kobayashi]{Philip~Taylor\thanks{E-mail: philip.1.taylor@anu.edu.au}$^{1,2}$ and Chiaki~Kobayashi$^2$\\
$^1$Research School of Astronomy \& Astrophysics, Australian National University, Mt. Stromlo Observatory, Cotter Rd., Weston, ACT 2611, Australia\\
$^2$Centre for Astrophysics Research, Science and Technology Research Institute, University of Hertfordshire, Hertfordshire, AL10 9AB, UK}

\begin{document}

\date{Accepted  Received ; in original form}

\pagerange{\pageref{firstpage}--\pageref{lastpage}} \pubyear{}

\maketitle

\label{firstpage}

\begin{abstract}

We predict the evolution of galaxy scaling relationships from cosmological, hydrodynamical simulations, that reproduce the scaling relations of present-day galaxies.
Although we do not assume co-evolution between galaxies and black holes a priori, we are able to reproduce the black hole mass--velocity dispersion relation.
This relation does not evolve, and black holes actually grow along the relation from significantly less massive seeds than have previously been used.
AGN feedback does not very much affect the chemical evolution of our galaxies.
In our predictions, the stellar mass--metallicity relation does not change its shape, but the metallicity significantly increases from $z\sim2$ to $z\sim1$, while the gas-phase mass-metallicity relation does change shape, having a steeper slope at higher redshifts ($z\ltsim3$).
Furthermore, AGN feedback is required to reproduce observations of the most massive galaxies at $z\ltsim1$, specifically their positions on the star formation main sequence and galaxy mass--size relation.


\end{abstract}



\section{Introduction}
\label{sec:4paper4_intro}

The formation and evolutionary histories of galaxies have been constrained from scaling relations.
In particular, the black hole mass--bulge mass relation indicates that black holes (BHs) and galaxies co-evolve, while the mass--metallicity relations suggest self-regulated chemical evolution of galaxies.
Large-scale surveys with ground-based telescopes, such as SDSS, have revealed these scaling relations, and with future space missions such as JWST these relations will be obtained at higher redshifts.
In this paper, we predict the time evolution of these relations in our cosmological simulations that include detailed chemical enrichment.

While the evolution of dark matter in the $\Lambda$ cold dark matter (CDM) cosmology is reasonably well understood, how galaxies form and evolve is much less certain because of the complexity of the baryon physics, such as star formation and feedback from supernovae and active galactic nuclei (AGN).
In galaxy-scale hydrodynamical simulations \citep[e.g.,][]{scannapieco12}, such small-scale physics has been modelled with some parameters, the values of which we have determined from observations of present-day galaxies \citep[][hereafter \citetalias{pt14}]{pt14}.
Although our modelling of star formation and stellar feedback is similar to other cosmological simulations \citep{vogelsberger14,schaye15,khandai15}, we have introduced a new AGN model where super-massive BHs originate from the first stars \citepalias{pt14}.
This AGN feedback results in better reproduction of the observed downsizing phenomena \citep[][hereafter \citetalias{pt15a}]{pt15a}, and cause AGN-driven metal-enhanced outflows from massive galaxies \citep[][hereafter \citetalias{pt15b}]{pt15b}.

The BH mass--bulge mass relation found by \citet{magorrian98} has underscored the importance of BHs during galaxy evolution.
The correlation with central velocity dispersion is one of the tightest and most straight forward to measure \citep{ferrarese00,gebhardt00,tremaine02,kormendyho13}.
Mass--metallicity relations \citep{faber73,kewley08, gallazzi08} contain information on how stars formed in galaxies, namely the importance of inflow and outflow during star formation.
This has been demonstrated with classical, one-zone models of chemical evolution \citep[e.g.,][]{tinsley80,matteucci01proc}, followed by hydrodynamical simulations \citep[e.g.,][]{ck07,finlator08}.
Colour--magnitude relations are the combination of the mass--metallicity relation and mass--age relation, although in early-type galaxies they are mostly caused by the metallicity effect \citep{kodama97}.
On stellar ages, the relation between specific star formation rates and stellar masses have been shown \citep{juneau05,stark13}, and the star formation main sequence \citep[e.g.,][]{renzini15} has become more popular in recent works.
The size--mass relation \citep{kormendy77,trujillo11} can also be reproduced with our cosmological simulations, which is important since this relation contains information on when stars form in collapsing dark halos, as demonstrated in simulations by \citet{ck05}.
The observed rapid evolution in size is one of the unsolved problems for early-type galaxies \citep{trujillo04,damjanov09}.

This paper is laid out as follows: in Section \ref{sec:4paper4_sims} we briefly describe the setup for our simulations, before showing presenting the evolution of various galaxy scaling relations from our simulations: the BH mass--velocity dispersion relation; mass--metallicity relations; size--mass relation; and the star formation main sequence, in Section \ref{sec:4paper4_res}.
Finally, we present our conclusions in Section \ref{sec:4paper4_conc}.
The box-averaged, cosmic evolution and stellar mass/luminosity functions are also presented in Appendix \ref{sec:4paper4_global} for reference.

\section{Simulations}
\label{sec:4paper4_sims}

Our simulations were introduced in \citetalias{pt15a}; they are a pair of cosmological, chemodynamical simulations, one run with the model for AGN feedback introduced in \citetalias{pt14}, the other identical but for its omission of AGN feedback.
Our simulation code is based on the {\sc GADGET-3} code \citep{springel05}, updated to include: star formation \citep{ck07}, energy feedback and chemical enrichment from supernovae \citep[SNe II\footnote{Our description of SNe II also includes a prescription for Type Ibc.} and Ia,][]{ck04,ck09} and hypernovae \citep{ck06,ck11}, and asymptotic giant branch (AGB) stars \citep{ck11a}; heating from a uniform, evolving UV background; and metallicity-dependent radiative gas cooling.
The BH physics that we include is described below.
We adopt the initial mass function (IMF) of stars from \citet{kroupa08} in the range $0.01-120\msun$, with an upper mass limit for core-collapse supernovae of $50\msun$.
These simulations have identical initial conditions consisting of $(240)^3$ particles of each of gas and dark matter in a periodic cubic box $25\,h^{-1}$ Mpc on a side, and show central clustering of galaxies.
This setup gives a gas particle mass of $M_{\rm gas}=1.44\times10^7\,h^{-1}\msun$, dark matter particle mass of $M_{\rm DM} = 7.34\times10^7\,h^{-1}\msun$, and gravitational softening length $\epsilon = 1.125\,h^{-1}$ kpc.
We employ a WMAP-9 $\Lambda$CDM cosmology \citep{wmap9} with $h=0.7$, $\Omega_{\rm m}=0.28$, $\Omega_\Lambda=0.72$, $\Omega_{\rm b}=0.046$, and $\sigma_8=0.82$.

A detailed description of our AGN model was presented in \citetalias{pt14}; we reiterate the key features here.
BHs are seeded from gas that is metal-free and denser than a threshold value in order to mimic the most likely channels of BH formation in the early Universe, specifically as the remnant of a massive stellar progenitor \citep[e.g.,][]{madau01,bromm02,schneider02,hirano14} or following the direct collapse of a massive, low angular momentum gas cloud \citep[e.g.,][]{loeb94,omukai01,bromm03,koushiappas04,agarwal12}.
We determined that a seed mass of $M_{\rm BH}=10^{2-3}\msun$ was necessary to best reproduce observations.
Such a seeding scheme is in contrast to other AGN feedback models \citep[e.g.,][]{vogelsberger14,schaye15} in which a single BH seed of mass $M_{\rm BH}\sim 10^5\msun$ is placed in every dark matter halo more massive than $\sim10^{10}\msun$.
Since our seeds are significantly less massive than other particles, we track the black hole mass due to accretion and mergers separately to the original particle mass.
BHs grow via Eddington-limited, Bondi-Hoyle gas accretion, and may also merge with other BHs if their separation is less than the gravitational softening length and their relative velocity is less than the local sound speed.
This velocity criterion ensures that BHs undergoing a flyby do not merge \citep[we note that the Illustris simulations do not use a velocity criterion; see][]{sijacki15}.
We note that the criterion $v_{\rm rel}<\sqrt{GM/\epsilon}$ has been used in other works \citep[e.g.,][]{booth09, schaye15}, where $M$ is the mass of the more massive of the merging BHs, and $\epsilon$ their gravitational softening.
We do not use this because the BH mass is calculated from its merger and accretion rates, whereas its velocity is determined using its `dynamical mass' $=\max(M_{\rm BH}, M_{\rm gas})$, which can be orders of magnitude larger than $M_{\rm BH}$.
The net effect of our AGN feedback is presented as the evolution of cosmic, box-averaged gas and stellar mass fractions and metallicities in Appendix \ref{sec:4paper4_global}.

In \citetalias{pt15a}, we showed how the final states of our simulations changed due to the action of AGN feedback by examining galaxy scaling relations, and found that it is essential in order to better match observed downsizing phenomena.
Despite the parameterization of small-scale physics due to the finite numerical resolution, our fiducial simulation gives good agreement with various physical properties of present-day galaxies.
We now extend this analysis to high redshift.
We compare the simulations to observations where available; all observationally derived galaxy masses are converted to a Kroupa IMF \citep{bernardi10}, unless a Chabrier IMF was used since they give very similar values \citep{bc03}.

In \citetalias{pt15a}, we introduced three galaxies, labelled A, B, and C, which we used to highlight the effects of AGN feedback on galaxies in different environments.
A is the most massive galaxy at the present day, and sits at the centre of a cluster, B is a massive field galaxy, and C is a low mass galaxy embedded in a dark matter filament away from the main cluster; more detailed descriptions, including star formation histories, may be found in \citetalias{pt15a}.
In section \ref{sec:4paper4_res}, we trace the positions of galaxies A, B, and C in the scaling relations back through the simulations.




\section{Results}
\label{sec:4paper4_res}

\subsection{$M_{\rm BH}$--$\sigma$}
\label{sec:4paper4_mbhsig}
\begin{figure}
\centering
\includegraphics[width=0.48\textwidth,keepaspectratio]{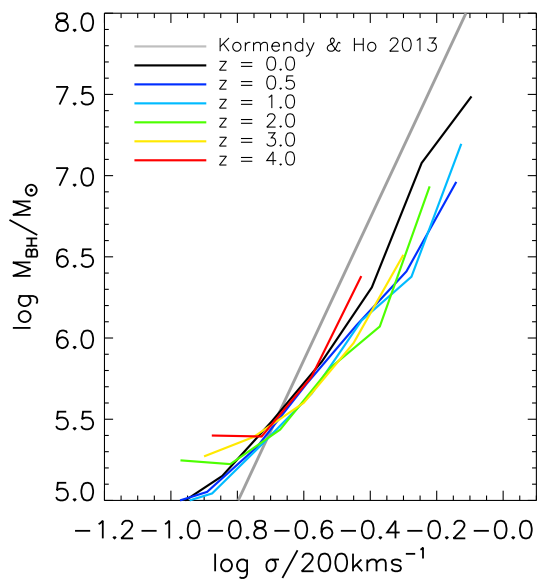}
\caption{Median $M_{\rm BH}$--$\sigma$ relation from our simulation at a range of redshifts.
The grey line shows the observed local relation from \citet{kormendyho13}.}
\label{fig:4paper4_mbhsig_med}
\end{figure}
\begin{figure*}
\centering
\includegraphics[width=\textwidth,keepaspectratio]{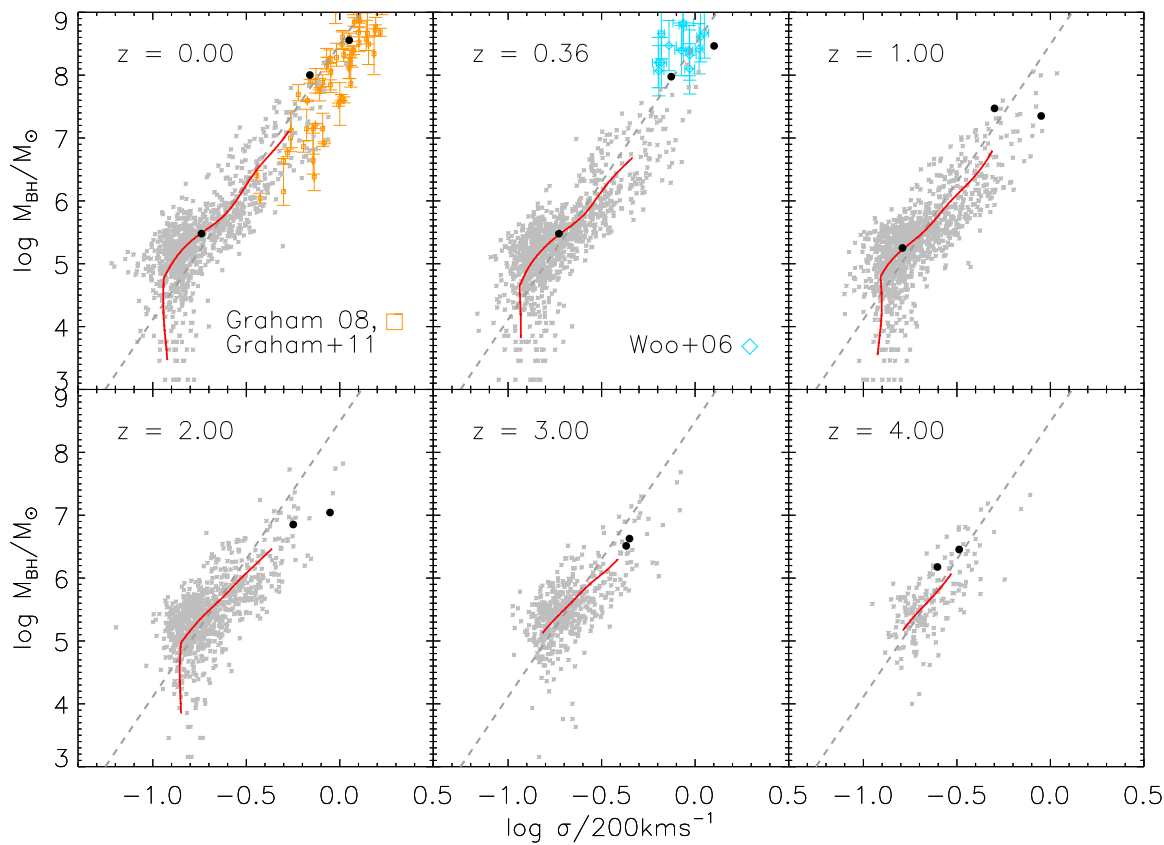}
\caption{Evolution of the $M_{\rm BH}$--$\sigma$ relation from our simulation.
In all panels, the red line shows the ridge line of the simulated relation, and the dashed grey line shows the observed local relation from \citet{kormendyho13}.
Also plotted are the observational data at $z=0$ from \citet{graham08,graham11}, and at $z=0.36$ from \citet{woo06}.
Black dots show the positions of galaxies A, B, and C.}
\label{fig:4paper4_mbhsig}
\end{figure*}

Of the many correlations between super-massive BH mass, $M_{\rm BH}$, and galaxy properties, the correlation with the central line-of-sight velocity dispersion of the stellar bulge, $\sigma$, is one of the tightest and most straightforward to measure.
Fig. \ref{fig:4paper4_mbhsig_med} shows the evolution of the median $M_{\rm BH}-\sigma$ relation at $z=0-4$ in our simulation with AGN, which shows no significant evolution.
$\sigma$ measured within the central projected $2h^{-1}$ kpc of a simulated galaxy.

In Fig. \ref{fig:4paper4_mbhsig}, we plot simulated $M_{\rm BH}$ against $\sigma$ at 6 redshifts ($z=0, 0.36, 1, 2, 3, 4$), comparing with observations.
In all panels, the grey dashed line corresponds to the local relation given in \citet{kormendyho13}, extrapolated below the observational limit of $\sim 10^6\msun$, and the solid red line denotes the ridge line of the 2D histogram of the data.
We are limited by resolution for $\log\left(\sigma/200 {\rm km\,s}^{-1}\right)\ltsim-0.8$, giving rise to the vertical feature present in most panels.
BHs therefore join the relation at this velocity dispersion, and $\log M_{\rm BH}/\msun=3-\log h$ (see \citetalias{pt14}), but quickly grow onto the observed relation.
While their mass is small, BH growth is dominated by mergers rather than gas accretion; this gives rise to the horizontal features that can be seen at $z\leq2$.
Our simulated $M_{\rm BH}$--$\sigma$ relation does not evolve, and lies on the observed local relation of \citet{kormendyho13} at all redshifts.
At $z=0$, the high-mass end of our relation is consistent with the data of \citet[][updated in \citealt{graham11}]{graham08}.
Data at $z>0$ are difficult to obtain due to the necessary resolution needed to obtain $M_{\rm BH}$ ($\sim1-10$ pc), but we include the measurements of \citet{woo06} at $z=0.36$, with which our data are also consistent.
The apparent offsets of $M_{\rm BH}--M_*$ often found in higher redshift observations \citep[e.g.,][]{peng06} are likely due to selection effects \citep{schulze14}.
A lack of or weak evolution in $M_{\rm BH}$--$\sigma$ (or the related $M_{\rm BH}$--$M_*$) has been seen in many other theoretical works, using different models and simulation techniques \citep[e.g.,][]{robertson06,dimatteo08,hirschmann14,degraf15,khandai15}.
This is due to the fundamental nature of the co-evolution of galaxies and BHs, which is not an a priori assumption in our model.

\begin{figure}
\centering
\includegraphics[width=0.48\textwidth,keepaspectratio]{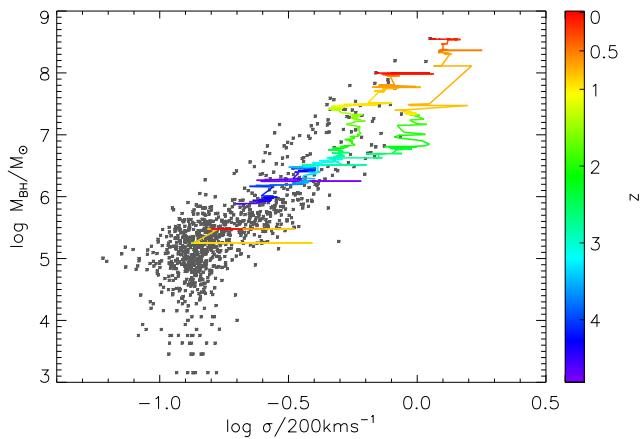}
\caption{Evolution of the $M_{\rm BH}$--$\sigma$ relation of galaxies A, B, and C from our simulation.
The position of each galaxy in the $M_{\rm BH}$--$\sigma$ plane is colour-coded by redshift.
Also shown is the $z=0$ relation for all galaxies in the simulation (grey stars).}
\label{fig:4paper4_mbhsigabc}
\end{figure}

The evolution of galaxies A, B, and C \citepalias{pt15a} is shown as well, with their positions on the $M_{\rm BH}$--$\sigma$ relation shown at each redshift by the black dots (galaxy A is the most massive of the three at each redshift).
To provide additional information for these three galaxies, we also show their $M_{\rm BH}$--$\sigma$ evolution with much finer time sampling in Fig. \ref{fig:4paper4_mbhsigabc}.
Galaxies A and B form at $z>4$, and lie close to the observed local relation at this time.
At $z\sim2.5$, A evolves away from the observed local relation, with $\sigma$ increasing more quickly than $M_{\rm BH}$, and grows back onto the relation at $z\sim0.5$.
Galaxy B, on the other hand, lies relatively close to the observed local relation at all times.
This difference is due to the different merger histories of the two galaxies; A undergoes several major mergers (including a triple merger at $z=0.75$) that increase its mass before gas is accreted by the BH, whereas the growth of B is dominated by more quiescent gas accretion that fuels both star formation and BH growth.
The narrow, horizontal features followed by rapid BH growth are caused by galaxy mergers followed by BH mergers.
Galaxy C forms between $z=1$ and $2$, and its BH grows quickly onto the observed local relation, but does not evolve at lower redshifts due to the passive nature of the galaxy's growth.
In summary, independent of the merging histories, BHs co-evolve with galaxies, which is indeed the origin of the tightness of the $M_{\rm BH}-\sigma$ relation.
Small deviation from the relation (i.e., smaller $M_{\rm BH}$ and larger $\sigma$) suggests that the galaxy has experienced a major merger just before the observed epoch.

\begin{figure*}
\centering
\begin{subfigure}{0.49\textwidth}
	\includegraphics[width=\textwidth,keepaspectratio]{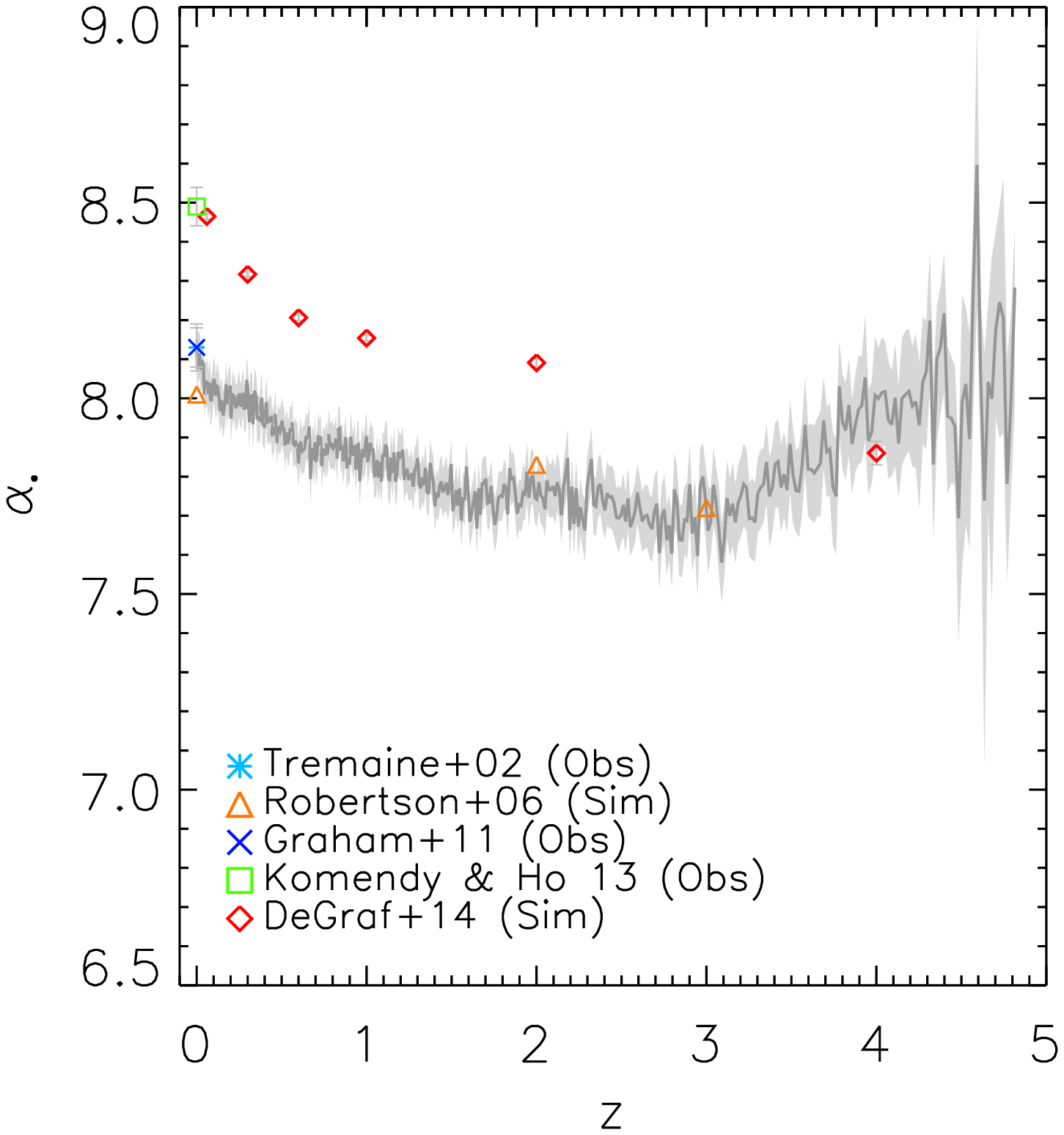}
	\caption{}
	\label{fig:4paper4_mbhsig_icpt}
\end{subfigure}
\begin{subfigure}{0.49\textwidth}
	\includegraphics[width=\textwidth,keepaspectratio]{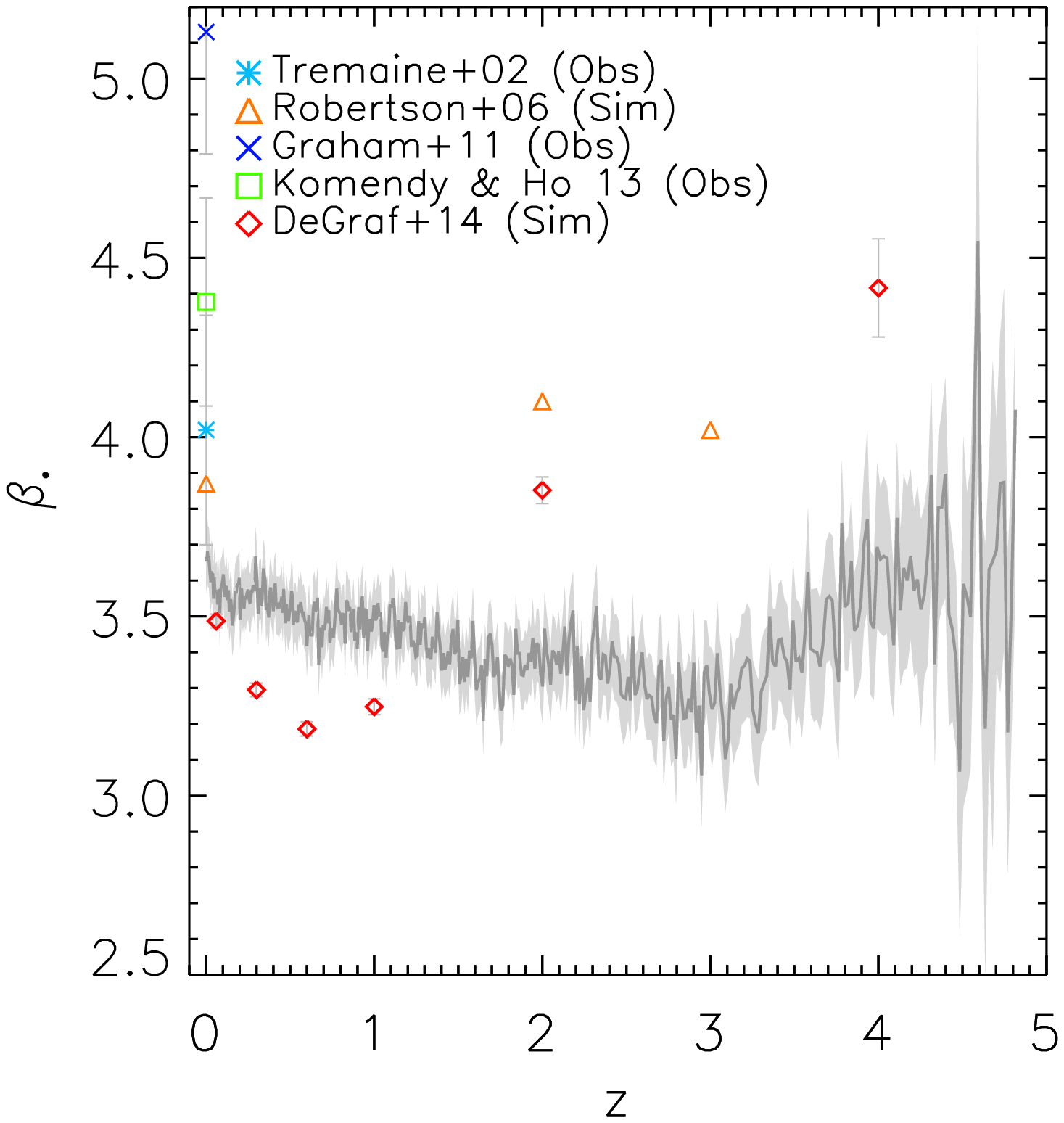}
	\caption{}
	\label{fig:4paper4_mbhsig_grad}
\end{subfigure}
\caption{Evolution of the fitted parameters of the $M_{\rm BH}$--$\sigma$ relation (see equation \eqref{eq:4paper4_mbhsig}).
The left-hand panel shows how the normalization, $\alpha_\bullet$, changes with redshift in our simulation (dark grey line and light grey shading), while the right-hand panel shows the evolution of the gradient, $\beta_\bullet$.
Also shown are the $z=0$ observations of \citet{tremaine02}, \citet{graham11}, and \citet{kormendyho13}, as well as the simulated evolution from \citet{robertson06} and \citet{degraf15}.}
\label{fig:4paper4_mbhsig_evol}
\end{figure*}

To provide a quantitative comparison with observations, we show in Fig. \ref{fig:4paper4_mbhsig_evol} the evolution of the intercept, $\alpha_\bullet$, and gradient, $\beta_\bullet$, of $M_{\rm BH}$--$\sigma$, defined by the equation
\begin{equation}\label{eq:4paper4_mbhsig}
\log\left(M_{\rm BH}/\msun\right) = \alpha_\bullet + \beta_\bullet\log\left(\sigma/200 {\rm km\,s}^{-1}\right).
\end{equation}
Most studies fit this relationship following the work of \citet[][see also \citealt{gultekin09,mcconnell13,graham16}]{tremaine02}, using a $\chi^2$ minimisation technique \citep[we also note that there are biases affecting the values of $\alpha_\bullet$ and $\beta_\bullet$ obtained from observations; see][]{shankar16}.
Due to the presence of the artificial features in our data discussed above, we employ a slightly different method.
We use least squares fitting, separately assuming $M_{\rm BH}$ and $\sigma$ is the independent variable, giving two fitted straight lines and two sets of parameters $\alpha_\bullet$ and $\beta_\bullet$.
Then our final best fit line is that which passes through the intersection of these lines and makes an equal angle with both (see Appendix \ref{app:lineaverage} for more details).
Errors on $\alpha_\bullet$ and $\beta_\bullet$ are obtained by repeating this procedure on 5000 bootstrap realisations of the simulated $M_{\rm BH}$--$\sigma$ distribution, and taking the standard deviation of the resulting $\alpha_\bullet$ and $\beta_\bullet$ distributions.
The left hand panel of Fig. \ref{fig:4paper4_mbhsig_evol} shows the evolution of $\alpha_\bullet$ with redshift for our simulations (dark grey line, light grey shaded area shows $1\sigma$ uncertainties).
At low redshift, our data are consistent with the observed values of \citet[][$\alpha_\bullet=8.13\pm0.06$]{tremaine02} and \citet[][$\alpha_\bullet=8.13\pm0.05$]{graham11} and the theoretical prediction of \citet{robertson06}, but lower than the observational value of \citet[][$\alpha_\bullet=8.490\pm0.049$]{kormendyho13} as well as \citet{degraf15}.
At higher redshift, the model predictions of \citet{robertson06} are consistent with our values.
Intriguingly, the evolution of $\alpha_\bullet$ in \citet{degraf15} at low redshift ($0<z<2$) shows a very similar trend to our data, but offset by $\sim 0.5$ dex.

The same is not seen in the right hand panel of Fig. \ref{fig:4paper4_mbhsig_evol}, where we show the evolution of $\beta_\bullet$ with redshift.
The theoretical values of \citet{robertson06} lie $\sim 0.5$ higher than ours at all $z$, while \citet{degraf15} show much stronger evolution with redshift than we predict.
The agreement with observations at $z=0$ is not so good as for $\alpha_\bullet$; the values of $4.02\pm0.32$ and $4.377\pm0.290$, from \citet{tremaine02} and \citet{kormendyho13} respectively, are $\sim 0.5-1$ larger than we measure, while the value of $5.13\pm0.34$ from \citet{graham11} is larger by almost 2.
It may be that the `bump' above the observed relation at $\log\left(\sigma/200{\rm km\,s}^{-1}\right)\sim-0.9$ is artificial, and caused by the very rapid growth of young BHs towards the relationship, but always at that velocity dispersion due to our finite resolution.
This feature may then cause the gradient to be underestimated, but our method to estimate $\beta_\bullet$ is meant to mitigate against this.
In any case, it is worth reiterating that the fitting procedure we used to obtain $\alpha_\bullet$ and $\beta_\bullet$ is different to that typically used in the literature, and that we have limited numbers of BHs with $M_{\rm BH}>10^6\msun$ because of the limited volume and finite seed mass.

\subsection{Mass--Metallicity Relation}
\label{sec:4paper4_massmet}
\begin{figure}
\centering
\includegraphics[width=0.48\textwidth,keepaspectratio]{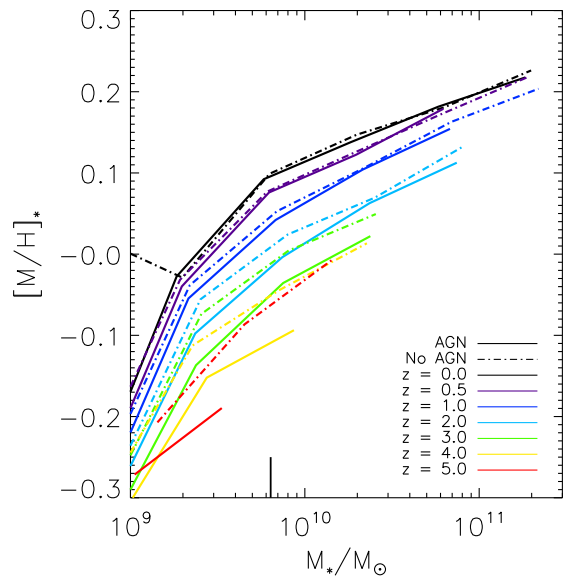}
\caption{The median $V$-band luminosity-weighted galaxy stellar MZR from our simulations at a range of redshifts.
Solid (dot-dashed) lines correspond to the simulation with (without) AGN feedback.
The short, vertical line indicates the mass of galaxies containing 1000 star particles.}
\label{fig:4paper4_zmass_med}
\end{figure}
\begin{figure*}
\centering
\includegraphics[width=\textwidth,keepaspectratio]{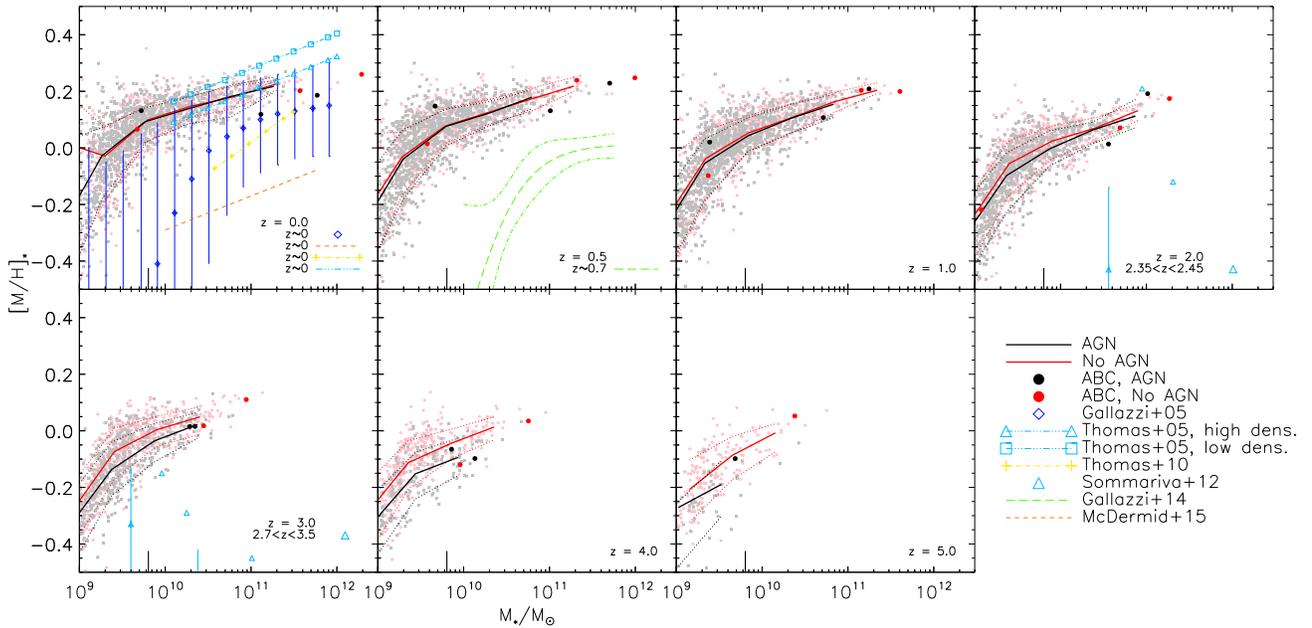}
\caption{Evolution of the luminosity-weighted galaxy stellar MZR.
Simulated data are shown by the black stars (with AGN) and red diamonds (without AGN), with their median (solid lines) and $1\sigma$ scatter (dotted lines) overlaid.
The median local relation and scatter from \citet{gallazzi05} are denoted by the blue diamonds and black error bars, and the fits to observations from \citet{thomas10}, \citet{sommariva12}, \citet{gallazzi14}, and \citet{mcdermid15} are also shown.
The short, vertical lines indicate the mass of galaxies containing 1000 star particles.}
\label{fig:4paper4_zmass}
\end{figure*}

Fig. \ref{fig:4paper4_zmass_med} shows the evolution of the stellar mass--stellar metallicity relation (MZR) at $z=0-5$ in our simulations with (solid lines) and without (dot-dashed lines) AGN.
Stellar metallicity, [M/H]$_*\equiv \log Z_*/Z_\odot$, is measured within the central projected $15$ kpc, weighted by $V$-band luminosity to compare with observations of absorption lines.
Due to the metallicity gradient present within galaxies, our results could be sensitive to this radius, but modest changes to it cause metallicities to change by less than the scatter in the MZR.
The lines denote the median relation of simulated galaxies.
There is only weak evolution in the stellar MZR at these redshifts, almost no evolution from $z=0$ to $z\sim 1$, and $\sim0.15$ dex increase from $z=4$ to $z=1$.
From $z=2-1$, the MZR experiences its strongest evolution as stars form from metal-rich gas after the peak of cosmic star formation.
With AGN, the shape of the MZR is very similar at all redshifts.
The slope is steeper at the low mass end, which is due to supernova feedback; more metals are ejected from low mass galaxies \citep{ck07}.

We show in Fig. \ref{fig:4paper4_zmass} the stellar MZR for our simulations, at various redshifts, and with observational data where available.
Values for individual galaxies in our simulations are shown by black stars (AGN) and red diamonds (no AGN), while the median and $1\sigma$ scatter are shown, respectively, by the solid and dotted lines of the same colours.
As shown in \citetalias{pt15a} and explained in \citetalias{pt15b}, AGN feedback has no significant effect on the MZR at this redshift, with both the median and scatter consistent between the simulations at all galaxy masses.
Only at high redshift ($z\geq4$) do the two simulations differ from one another; the galaxies in the simulation with AGN have metallicities on average $\sim0.1$ dex lower than their counterparts in the other simulation, comparable to the $1\sigma$ scatter, which is due to the delayed onset of SF due to AGN feedback.

At $z=0$, both simulations agree well with the fits of \citet[][cyan squares and triangles]{thomas05}, especially for $10^{10}\leq M_*/\msun\leq 2\times 10^{11}$, although the most massive galaxies are under-enriched by $\sim 0.1$ dex with respect to these observations.
The SDSS sample (\citealt{gallazzi05}, blue diamonds; \citealt{thomas10}, yellow ticks) show a very similar slope, but with $\sim 0.1$ dex lower metallicities.
This difference may be due to the different analysis of the observational data.
The fit of \citet{mcdermid15} to data from the ATLAS$^{\rm 3D}$ survey shares the gradient found by \citet{thomas05} using a relatively small sample of early-type galaxies, but is offset by $\sim-0.4$ dex, the reason for which is uncertain \citep{mcdermid15}.
\citet{gallazzi05} shows a rapid decrease at the low mass end, which is weakened by the analysis of \citet{panter08}.
At $z>0$, very few observational data are available; we show the $z\sim0.7$ MZR of \citet{gallazzi14} alongside our $z=0.5$ results.
Our simulated MZR shows enrichment at least $0.2-0.3$ dex higher than observed.
At higher redshift, we include the observational data from \citet[][cyan triangles]{sommariva12}, which tend to lie below our predictions, though with a large amount of scatter.
Estimating stellar metallicity is difficult to do with observations at high redshift due to the spatial resolution, and even at $z=0$ because of the age-metallicity degeneracy, and so there can be a significant offset between different authors, as well as large scatter in individual catalogues.
For example, the $1\sigma$ scatter in the relation for SDSS galaxies at $z=0$ from \citet[][blue diamonds and error bars in top left panel]{gallazzi05} is larger than the $1\sigma$ scatter from our simulations, shown by the dotted lines.
It also lies offset from our median simulated relations, with a greater difference at lower masses.

The evolution of the position of galaxies A, B, and C in the MZR is also shown by the black and red dots (with and without AGN, respectively; A, B, and C are ordered by mass at all redshifts in both simulations).
All three galaxies grow along the relation, which explains the low scatter of this relation in our simulations, and by $z\sim1-2$, A and B have attained their present-day metallicities, while C continues to be enriched to low redshift as star formation is not totally quenched.

\begin{figure}
\centering
\includegraphics[width=0.48\textwidth,keepaspectratio]{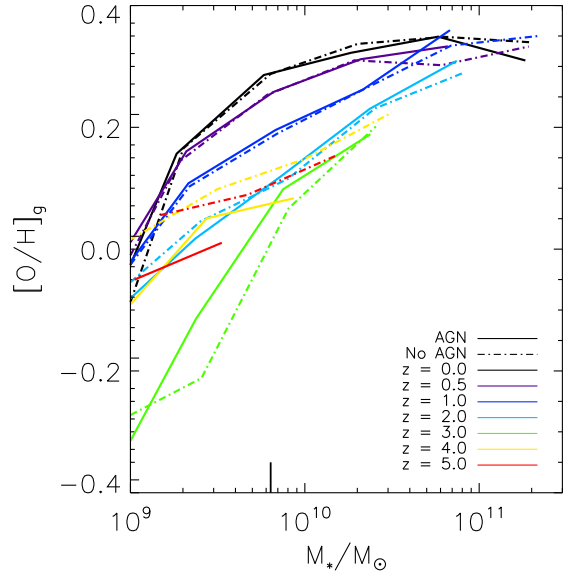}
\caption{The median SFR-weighted galaxy gas-phase MZR from our simulations at a range of redshifts.
Solid (dot-dashed) lines correspond to the simulation with (without) AGN feedback.
The short, vertical line indicates the mass of galaxies containing 1000 star particles.}
\label{fig:4paper4_omass_med}
\end{figure}
\begin{figure*}
\centering
\includegraphics[width=\textwidth,keepaspectratio]{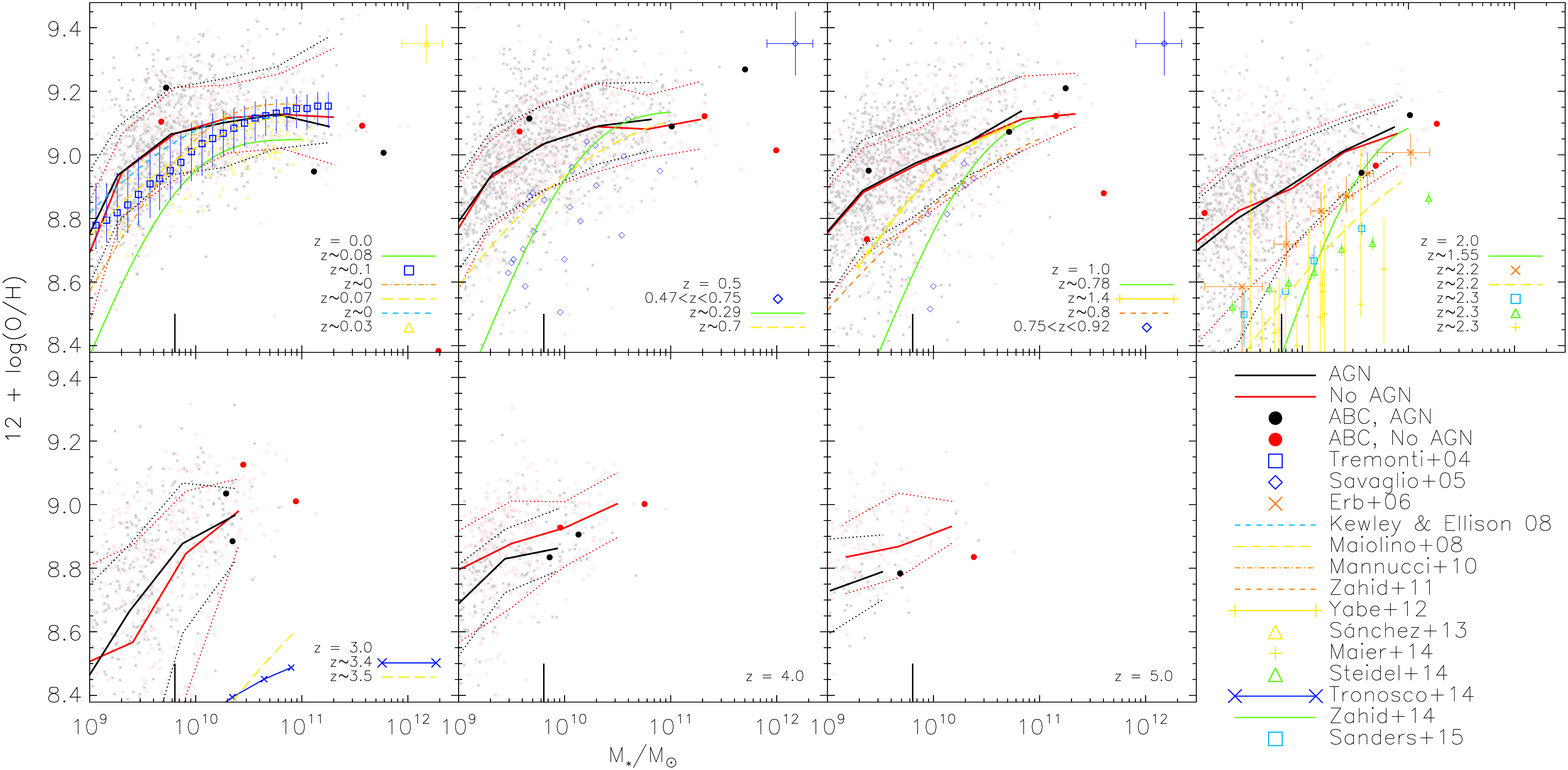}
\caption{Evolution of the SFR-weighted galaxy gas-phase MZR.
Simulated data are shown by the black stars (with AGN) and red diamonds (without AGN), with their median (solid lines) and $1\sigma$ scatter (dotted lines) overlaid.
Also shown are the observational data of \citet{tremonti04}, \citet{savaglio05}, \citet{erb06}, \citet{sanchez13}, \citet{maier14}, \citet{steidel14}, and \citet{sanders15}, as well as the fits to data from \citet{kewley08}, \citet{maiolino08}, \citet{zahid11}, \citet{yabe12}, \citet{tronosco14}, and \citet{zahid14}.
All observational data have been converted to our metallicity scale, a Kroupa IMF, and to the method of \citet{kewley02} using the procedure given in \citet{kewley08}.
The short, vertical lines indicate the mass of galaxies containing 1000 star particles.}
\label{fig:4paper4_omass}
\end{figure*}

Fig. \ref{fig:4paper4_omass_med} shows the evolution of the gas-phase MZR at $z=0-5$ in our simulations with (solid lines) and without (dot-dashed lines) AGN.
We measure gas oxygen abundance within the central $15$ kpc of each galaxy (regardless of the gas' temperature), weighted by SFR to compare with observations, which are estimated from emission lines.
The lines denote the median of the simulated relation.
There is much greater evolution in the gas-phase MZR at these redshifts than the stellar MZR.
The evolution at $z=0-0.5$ is weak, but the metallicities systematically decrease until $z=3$, with a steeper slope at high redshifts.
However, the relation becomes flatter from $z=3-4$, with fewer low metallicity galaxies at a given mass at higher redshift.
This is because low mass galaxies that form earlier have rapid star formation enriching their ISM, while later galaxies form stars more slowly with larger gas accretion,which is evidence for downsizing in our simulations.

The simulated results are shown in Fig. \ref{fig:4paper4_omass} displayed in the same way as in Fig. \ref{fig:4paper4_zmass}.
Unlike the stellar MZR, the gas-phase MZR is consistent between the two simulations to high redshift.
Galaxies A and B reach their present-day metallicities at $z\sim1-2$ (though low-redshift AGN-driven winds subsequently remove much of the gas in these galaxies), while for galaxy C this happens at lower redshift (and more quickly with AGN feedback since enriched winds from the central cluster help to pollute this galaxy).

All observational data from the literature have been converted to a Kroupa IMF (unless a Chabrier IMF was used, since this gives very similar results), converted to our adopted solar metallicity, $12+\log({\rm O}/{\rm H})_\odot=8.78$, and converted to the method of \citet{kewley02} using the procedure given in \citet{kewley08}.
In cases where the assumed solar value was not given, we adopted $12+\log({\rm O}/{\rm H})_\odot=8.66$ \citep{asplund05} for papers earlier than 2009, and $12+\log({\rm O}/{\rm H})_\odot=8.69$ \citep{asplund09} otherwise.
At $z=0$, in the top left panel of Fig. \ref{fig:4paper4_zmass}, there is excellent agreement between our simulations (which are fully consistent with one another) and the observations of \citet{tremonti04}, \citet{kewley08}, \citet{maiolino08}, and \citet{sanchez13} at all galaxy masses, as well as \citet{mannucci10} at $M_*>10^{10}\msun$.
The fit of \citet[][solid green line]{zahid14} is systematically lower than both our simulations and other observational data by $\sim0.1$ dex, the reason for which is unknown.
We also find the scatter in the simulated MZRs to be greater than observed.
This may be because we do not consider only cold ($T<1.5\times10^4$ K) gas, which observations are sensitive to, though weighting by SFR should alleviate this effect.
At higher redshifts, out to $z=2$, our simulations tend to agree well with observational data for high-mass galaxies, but show excess enrichment in low-mass galaxies compared to observations.
By $z=3$, where few data are available, the simulated relations lie $\sim0.5$ dex above the observations of \citet{maiolino08} and \citet{tronosco14} at all masses.

\subsection{Size--Mass Relation}
\begin{figure}
\centering
\includegraphics[width=0.48\textwidth,keepaspectratio]{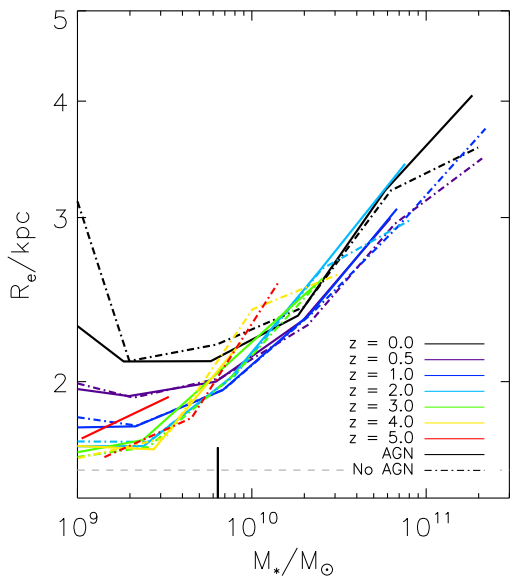}
\caption{The galaxy size--mass relation of our simulations at a range of redshifts.
Solid (dashed) lines correspond to the simulation with (without) AGN feedback.
The short, vertical line indicates the mass of galaxies containing 1000 star particles, while the dashed horizontal line shows the gravitational softening length.}
\label{fig:4paper4_rem_med}
\end{figure}
\begin{figure*}
\centering
\includegraphics[width=\textwidth,keepaspectratio]{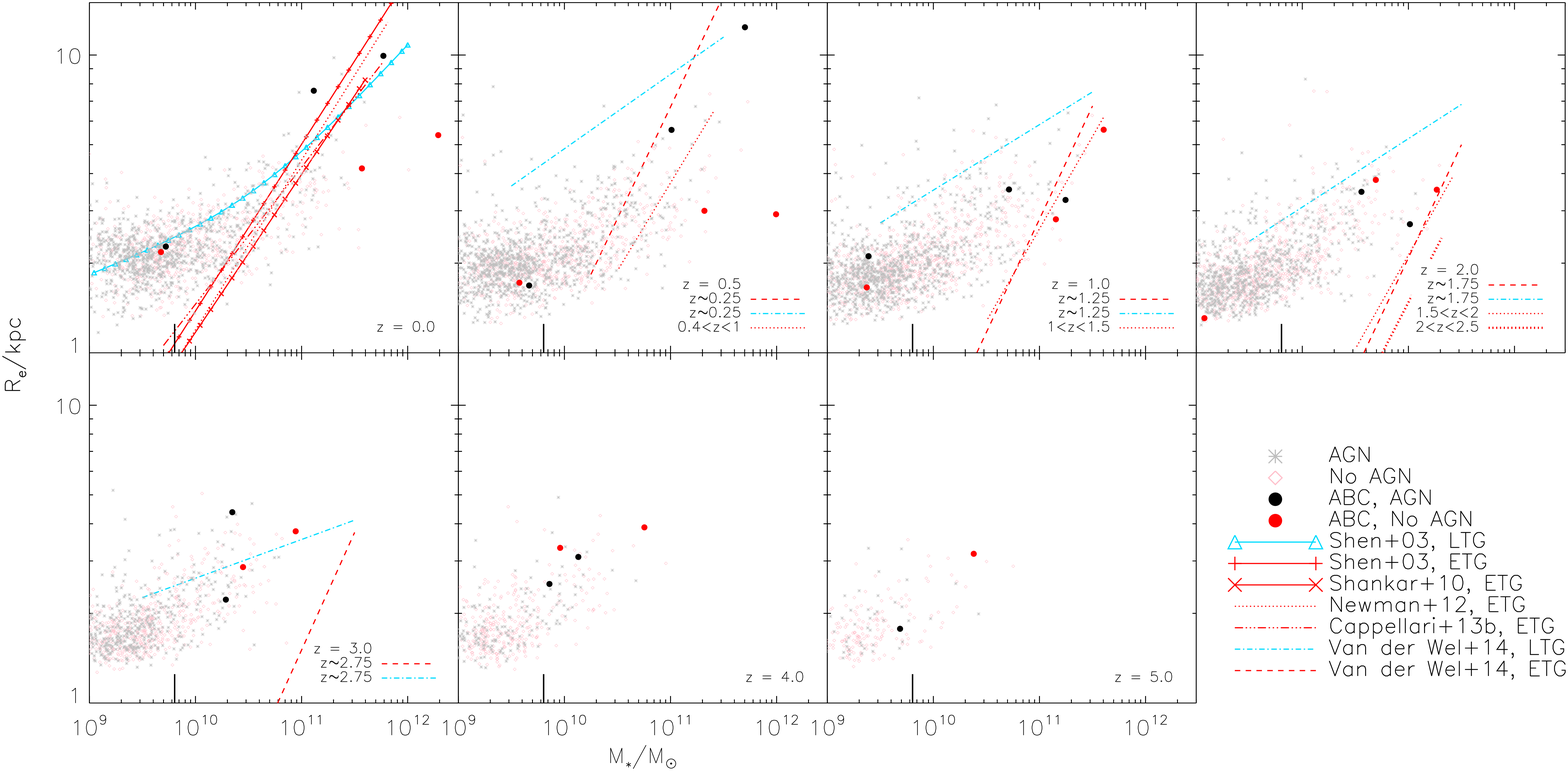}
\caption{Evolution of the galaxy size--mass relation.
Simulated data are shown by the black stars (with AGN) and red diamonds (without AGN).
Fits to observed relations are shown for early-type galaxies by the red lines \citep{shen03,shankar10,newman12,cappellari13,vanderwel14}, and for late-type galaxies by the blue lines \citep{shen03,vanderwel14}.
The short, vertical lines indicate the mass of galaxies containing 1000 star particles.}
\label{fig:4paper4_rem}
\end{figure*}

The size--mass relation of galaxies gives information about where, on average, stars formed in the initial collapsing gas cloud, and therefore can constrain the SF timescale \citep{ck05}.
However, cosmological simulations have struggled to produce galaxies as large as observed, and AGN feedback has been found to offer a solution by suppressing star formation at the centres of massive galaxies \citep[e.g.,][but see also \citealt{snyder15}'s analysis of galaxies in the Illustris simulation, which are larger than observed]{dubois13,pt14}.
At high redshifts, it has been shown that spheroid-like galaxies are much more compact than at present \citep[e.g.,][]{trujillo04}.
The favoured formation scenario for this size evolution is not major mergers or puffing up by AGN of stellar winds, but minor mergers \citep[see][for a review]{trujillo13}.
In our cosmological simulations, it is very hard to determine the morphology of galaxies due to the limited resolution, and we show the size evolution simply as a function of stellar mass.
Note that at high redshifts, the fraction of early-type massive galaxies decreases, and large galaxies may have young stellar populations \citep{belli15}.

Effective radii are calculated differently from in \citetalias{pt14} and \citetalias{pt15a}, where they were defined as the circular radius enclosing half of the galaxy's stellar mass.
In order to compare more consistently with observations, we fit elliptical isophotes to the 2D bolometric luminosity distribution of each galaxy, and fit a core-S\'ersic function \citep{graham03,trujillo04} to the intensity profile as a function of $r=\sqrt{ab}$, where $a$ and $b$ are the semi-major and semi-minor axis lengths, respectively.
For galaxies with a luminosity distribution that is not well fit by ellipses (which tend to be merging or poorly resolved systems), the intensity profile is obtained using circular annuli.
The effective radius is then the value of $r$ that encloses half the total light from the galaxy.
Some galaxies still have a bad fit, and are not included in the following figures.
Galaxy stellar masses are from a friends-of-friends code used to identify galaxies, as in \citetalias{pt14} and \citetalias{pt15a}.

The median size--mass relation for both simulations with and without AGN at several redshifts is shown in Fig. \ref{fig:4paper4_rem_med}.
In both simulations, the galaxy sizes become smaller at higher redshifts, at the low mass range.
The difference between the simulations at $z\ltsim0.5$ is not discernible due to the small number of galaxies that are strongly affected by AGN.
This suggests that the primary process of the size evolution is not related to AGN, but to minor mergers that occur naturally in cosmological simulations.
We should note that the relationship flattens towards lower masses at $R_{\rm e}\sim1-2$ kpc, which is comparable to the gravitational softening length ($1.125\,h^{-1}$ kpc at $z=0$).

The size evolution in the simulations is not as large as observed.
In Fig. \ref{fig:4paper4_rem}, we show the galaxy size--mass relation for our simulated galaxies at various redshifts, as well as several published observed relations (blue lines are for late-type galaxies, red for early-type galaxies).
The effects of AGN feedback are only apparent in the most massive galaxies, but at $z=0$ this steepens the relationship, and brings it more in line with observations \citep{shen03,shankar10,newman12,cappellari13}, i.e. the time evolution is stronger with AGN feedback.
At higher redshift, the difference between the simulations with and without AGN becomes less pronounced because BHs have had less time to influence star formation in galaxies since the peak of cosmic star formation at $z=2$, which is almost unaffected by AGN feedback.
There is, however, between $z\sim1$ and 3, an excess of galaxies with AGN that lie above the main relation and close to the observed relation for late-type galaxies of \citet{vanderwel14} compared to the simulation without AGN.
Other than their large radii, these galaxies do not exhibit properties that set them apart; their colours are not systematically bluer, as might be expected for late-type galaxies, and they may be due to imperfect fits.

The three example galaxies, A, B, and C, are denoted by the black and red dots (AGN and no AGN, respectively).
There is little difference in their evolution in the different simulations until $z<1$, when A and B exhibit much larger effective radii in the simulation with AGN feedback, due to the suppression of central star formation at late times.
Galaxy C, on the other hand, evolves consistently in both simulations since it does not experience strong AGN feedback at any time.

\subsection{Star Formation Main Sequence}
\label{sec:4paper4_sfms}
\begin{figure}
\centering
\includegraphics[width=0.48\textwidth,keepaspectratio]{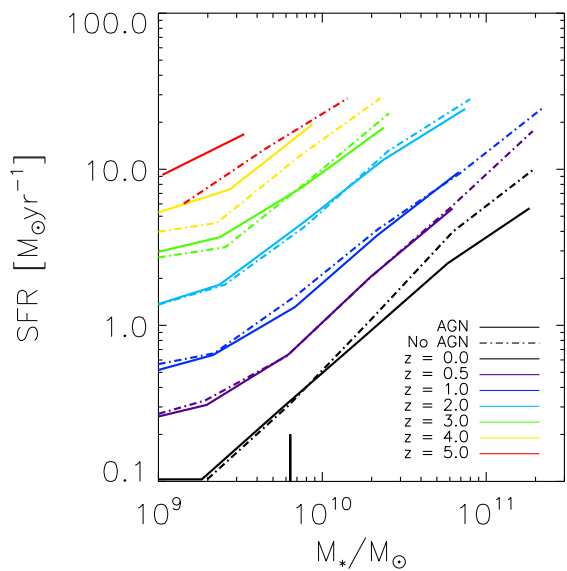}
\caption{The SFMS in our simulations at a range of redshifts.
Solid (dashed) lines correspond to the simulation with (without) AGN feedback.
The short, vertical line indicates the mass of galaxies containing 1000 star particles.}
\label{fig:4paper4_sfms_med}
\end{figure}
\begin{figure*}
\centering
\includegraphics[width=\textwidth,keepaspectratio]{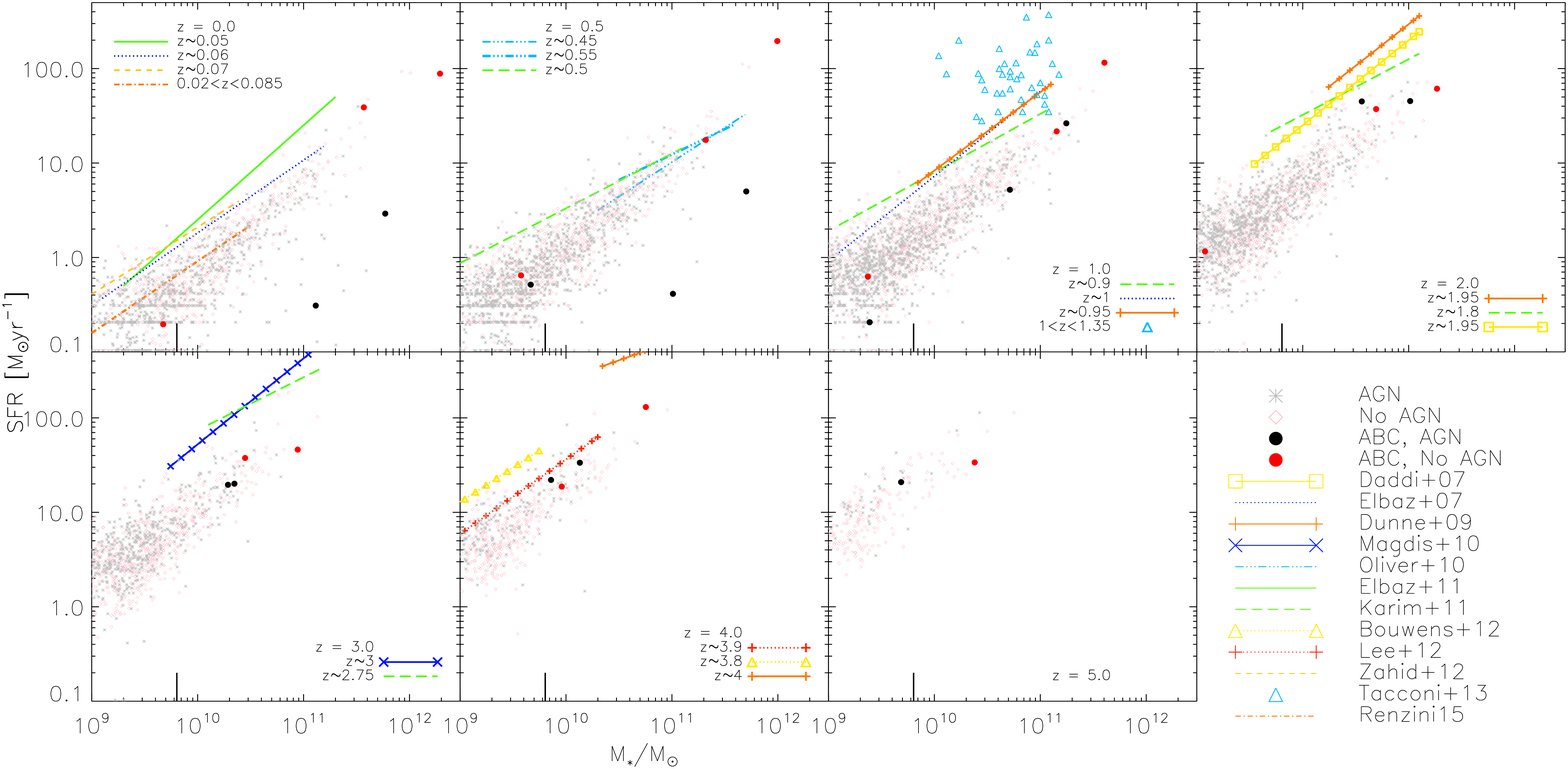}
\caption{Evolution of the SFMS.
Simulated data are shown by the black stars (with AGN) and red diamonds (without AGN).
Observational data from \citet{tacconi13} are also plotted, along with the fits to observations of \citet{daddi07}, \citet{elbaz07}, \citet{dunne09}, \citet{magdis10}, \citet{oliver10}, \citet{elbaz11}, \citet{karim11}, \citet{bouwens12}, \citet{lee12}, \citet{zahid12}, and \citet{renzini15}.
The short, vertical lines indicate the mass of galaxies containing 1000 star particles.}
\label{fig:4paper4_sfms}
\end{figure*}

The star formation main sequence (SFMS) describes the correlation between galaxy mass and SFR, with more massive galaxies tending to have larger SFRs.
The simulated SFRs are found from the total of the initial stellar masses of star particles that formed in the preceding $10^8$ yr, divided by $10^8$ yr.
Not surprisingly, more massive galaxies tend to have higher SFRs.
In Fig. \ref{fig:4paper4_sfms_med}, we show the median SFMS for both of the simulations with and without AGN at different redshifts, which indicates that AGN feedback has very little effect on it except at the high-mass end.
In addition, while the normalisation shows quite strong evolution with redshift, the gradient does not change, which is consistent with observations \citep{whitaker14}.

Fig. \ref{fig:4paper4_sfms} shows the SFMS for our simulated galaxies at various redshifts, along with fits to observational data \citep{daddi07,elbaz07,dunne09,magdis10,oliver10,elbaz11,karim11,bouwens11,lee12,zahid12,tacconi13,renzini15}.
AGN feedback is particularly important for galaxies that lie in the region of the $M_*$--SFR plane occupied by early-type galaxies \citep[e.g.,][]{wuyts11,renzini15}.
In our simulated galaxies, AGN feedback is quenching star formation.
Therefore there is a clear dearth of high-mass ($M_*>10^{11}\msun$) galaxies on the SFMS compared to the simulation without AGN, but it is otherwise unchanged by the inclusion of AGN feedback.
In terms of evolution, our simulations are consistent with observations, with a very similar gradient, but slightly lower normalisation.
At $z=0$ in particular, fitting a straight line to our simulated SFMS gives $\log {\rm SFR} = (-0.11\pm0.01)+(0.77\pm0.03)\log M_*/10^{10}$ in excellent agreement with the fit of \citet{renzini15} to SDSS galaxies: $\log {\rm SFR} = (-7.64\pm0.02)+(0.76\pm0.01)\log M_*$.
At $z\gtsim2$, the offset between observations and simulations becomes greater than the scatter in our observed relations at all masses.
This is surprising since our simulation is consistent with, or even slightly higher than, the observed cosmic SFR \citepalias[see Fig. 2 of][]{pt15a}.
At these high redshifts, observations may suffer from a Malmquist bias.
Additionally, star formation tracers from observations typically probe timescales $<10^8$ yr, which may contribute to the offset seen before the peak of star formation at $z\sim2$.
It is worth noting that with recent estimates of dust obscured star formation, the evolution in SFMS is found to be weaker, in excellent agreement with our predictions (Smith, D. J. B., et al. 2016, MNRAS, submitted).

Galaxies A, B, and C are shown as the black and red dots (with and without AGN, respectively).
These galaxies in the simulation with AGN tend to have a lower SFR than their counterparts in the other simulation.
This is most clearly seen for galaxies A and B at $z\ltsim0.5$, when these galaxies move off the SFMS, due to the suppression of star formation, and into a region occupied, observationally, by early-type galaxies \citep{wuyts11}.

\begin{figure}
\centering
\includegraphics[width=0.48\textwidth,keepaspectratio]{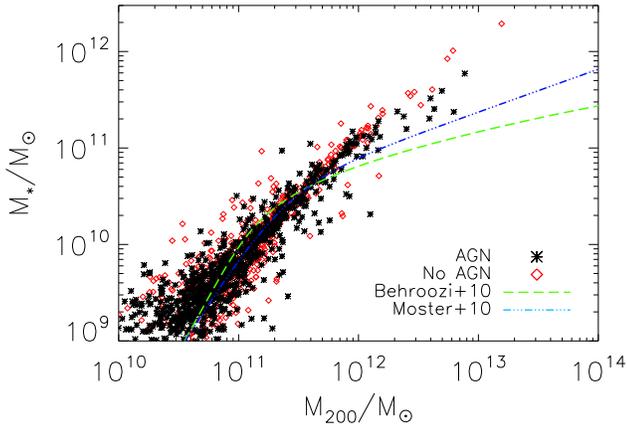}
\caption{Stellar mass--halo mass relation at $z=0$.
Simulated data are shown by the black stars (with AGN) and red diamonds (without AGN).
Fits to observational data from \citet{behroozi10,moster10} are also shown (green dashed and blue dot-dashed lines, respectively), with a shift for the different definition of halo masses (see the text for the details).}
\label{fig:smhm}
\end{figure}
Related to the SFMS, we show in Fig. \ref{fig:smhm} the stellar mass -- halo mass relation from our simulations at $z=0$, as well as fits to observational data \citep{behroozi10,moster10}.
Note that these observations used the virial halo masses, while we show $M_{200}$, which may cause some offset in halo mass with respect to our simulated data.
We have therefore plotted their relations with a shift of $-0.54$\,dex applied to their halo masses to facilitate comparison.
Although our simulation box do not cover the mass range of the observations, the influence of AGN feedback at $\sim3\times10^{11}\msun$ is clear, reducing their stellar mass at given halo mass.
Even stronger feedback may be required as discussed in \citet{pt15a}.

\section{Conclusions}
\label{sec:4paper4_conc}

Using our cosmological, chemo-dynamical simulation that well reproduces the observed scaling relations of present-day galaxies, we predict the redshift evolution of the scaling relations.
In keeping with previous theoretical works, we find no evolution of the $M_{\rm BH}$--$\sigma$ relation at $z<4$, and show that BHs grow along the relation despite the relatively low seed mass used in our model (Fig. \ref{fig:4paper4_mbhsigabc}).
This is due to the fundamental nature of the co-evolution of galaxies and BHs, which is not an a priori assumption in our AGN model.
Furthermore, we predict the intercept and gradient of the $M_{\rm BH}$--$\sigma$ relation as $\alpha_\bullet\approx8$ and $\beta_\bullet\approx3.5$, respectively.
This value for the intercept is consistent with both observational estimates at $z=0$ and the results of other simulations at all redshifts.
Our estimate for the gradient is lower than observational estimates in particular, though there is evidence that these may be overestimated \citep{shankar16}.

The influence of AGN feedback on galaxy formation and evolution, highlighted through comparison of our two simulations with and without AGN feedback, is most apparent at $z<2$, after the peak of cosmic SFR, and preferentially in more massive galaxies.
These galaxies host the most massive BHs (Fig. \ref{fig:4paper4_mbhsig}), and may have high accretion rates at late times when they have grown to be super-massive.
The primary effect of this feedback is to quench star formation in the central regions of massive galaxies, thereby reducing the prevalence of high-mass and high-luminosity galaxies, increasing their effective radii ($z<1$; Fig. \ref{fig:4paper4_rem}), and moving them off the star formation main sequence at low redshift ($z\ltsim0.5$; Fig. \ref{fig:4paper4_sfms}).

However, the chemical evolution of galaxies is not much affected by AGN, and the MZRs (Figs. \ref{fig:4paper4_zmass} and \ref{fig:4paper4_omass}) are mostly established at a galaxy's first starburst, which is typically before its BH has grown sufficiently to strongly affect star formation and stellar feedback.
In our predictions, the stellar MZR does not change shape, but the normalization changes significantly from $z\sim2$ to $z\sim1$ as stars form from metal-rich gas after the peak of cosmic star formation.
The gas-phase MZR does change shape, with a steeper slope at high redshifts.
Observations of these relations at high redshift with instruments such as JWST are of particular importance to constrain the formation processes of the majority of the stellar population in galaxies.

We also compare our simulations to available observations, finding excellent agreement in many cases at low redshift, and fairly good agreement out to $z=2-3$, especially for the most massive galaxies, with the agreement usually better for the simulation with AGN feedback (for the size--mass relation in particular, Fig. \ref{fig:4paper4_rem}).
Our size evolution is driven mainly by minor mergers that occur naturally in cosmological simulations, and is not as large as observed even with AGN feedback.
We also find that the simulated mass and $K$-band luminosity functions (see Appendix \ref{sec:4paper4_mfn}) significantly overpredict the number of low-mass galaxies at all redshifts, and that their evolution is not in keeping with observations.
A potentially related issue is that low-mass galaxies in our simulations are over-enriched compared to observations (Fig. \ref{fig:4paper4_omass}).
This may be due, at least in part, to our initial condition.
It may also be the case that baryon physics is more complicated than in our prescriptions.

\section*{Acknowledgements}

PT acknowledges funding from an STFC studentship.
This work was supported by a Discovery Projects grant from the Australian Research Council (grant DP150104329).
This work has made use of the University of Hertfordshire Science and Technology Research Institute high-performance computing facility.
This research was undertaken with the assistance of resources provided at the NCI National Facility systems at the Australian National University through the National Computational Merit Allocation Scheme supported by the Australian Government.
This research made use of the DiRAC HPC cluster at Durham.
DiRAC is the UK HPC facility for particle physics, astrophysics, and cosmology, and is supported by STFC and BIS.
CK acknowledges PRACE for awarding her access to resource ARCHER based in the UK at Edinburgh.
Finally, we thank V. Springel for providing {\sc GADGET-3}.


\bibliographystyle{mn2e}
\bibliography{./refs}


\appendix

\section{Evolution of Cosmic Properties}
\label{sec:4paper4_global}

\begin{figure*}
\centering
\includegraphics[width=\textwidth,keepaspectratio]{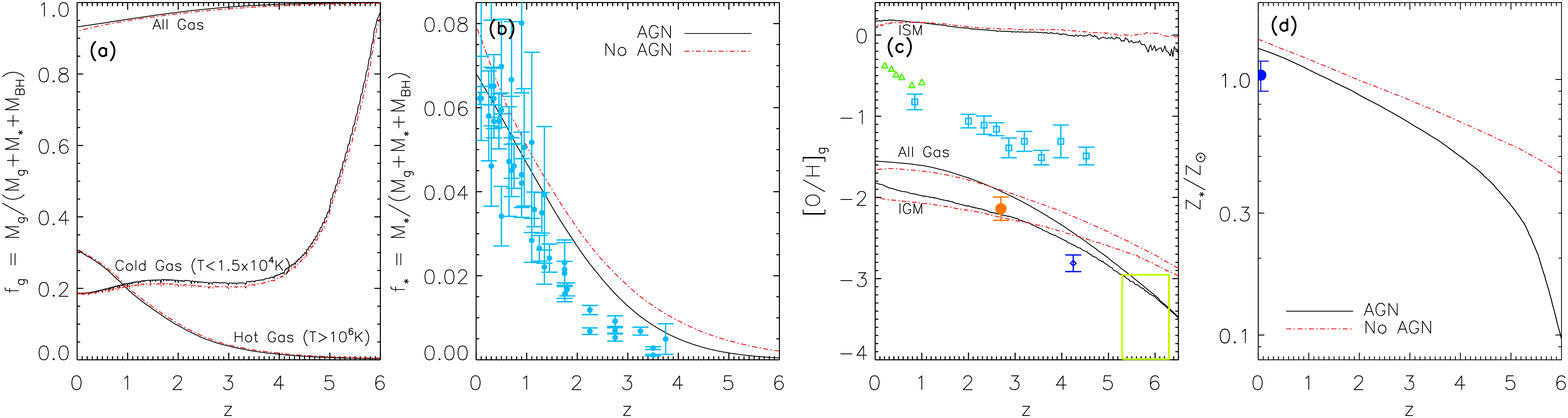}
\caption{{\bf a}) Cosmic gas fraction $M_{\rm g}/(M_{\rm g}+M_{\rm *}+M_{\rm BH})$ as a function of redshift for all gas (left panel), hot gas ($T>10^6$K), and cold gas ($T<1.5\times10^4$K).
{\bf b}) Cosmic stellar fraction $M_*/(M_{\rm g} + M_* + M_{\rm BH})$; observational data (cyan circles) are taken from the compilation of stellar densities in \citet{madau14}.
{\bf c}) Gas-phase oxygen abundances as a function of redshift for all gas, ISM, and IGM.
Observational data are from: \citet[][green triangles]{balestra07}; \citet[][cyan squares]{rafelski12}; \citet[][orange dot]{aguirre08}; \citet[][blue diamond]{simcoe11}; \citet[][lime box]{ryanweber09,simcoe11,becker11}.
{\bf d}) Cosmic mass-weighted stellar metallicity as a function of redshift; observational data (blue circle) is taken from \citet{gallazzi08}.
In all panels, solid black lines are for the simulation with AGN, and the red dot-dashed lines are for the simulation without AGN.}
\label{fig:4paper4_fg_all}
\end{figure*}

Here we show the next effect of our AGN feedback in the box-averaged, cosmic evolutions.
Needless to say, chemical enrichment takes place inhomogeneously, highly depending on the environment, in the Universe.
This should be considered in the comparison to the observations of cosmic quantities.

Cosmic star formation rates (SFR) are strongly affected by our AGN feedback, with lower rates at both low and high redshifts \citepalias[Fig. 2 of][]{pt15a,pt14}.
However, individual BHs in our simulations directly affect only their local environments, with even the most massive affecting a region $\sim 1$ Mpc across \citepalias{pt15b}.
It is therefore interesting to examine how other cosmic quantities of the simulated gas and stars change.
To this end, we show the cosmic gas fraction, defined as $f_{\rm g} = M_{\rm g}/(M_{\rm g}+M_{\rm *}+M_{\rm BH})$, as a function of redshift in Fig. \ref{fig:4paper4_fg_all}a.
The total gas fraction falls from $f_{\rm g}=1$ at high redshift, to $f_{\rm g}\sim0.93$, which compares well with the observational estimate of \citet{fukugita04}, $f_{\rm g}=0.95\pm0.09$.
The cold gas fraction also shows a decreasing trend with time, while the fraction of hot gas increases, especially at $z\ltsim2$ when large scale structure has collapsed and feedback from stars and AGN is most effective.
In all cases, the effects of AGN appear to be very small in this figure.
In particular, the hot gas fraction is almost the same, which is due to the self regulation; the simulation with AGN has lower SFRs and thus less supernova feedback.
The largest difference is seen at intermediate redshift ($1\ltsim z \ltsim 3$) in cold gas, with the simulation with AGN having the larger cold gas fraction, which may be due to the reduced rate of star formation at higher redshift compared to the other simulation.

Figure \ref{fig:4paper4_fg_all}b shows the cosmic stellar fraction, $f_{\rm *} = M_{\rm *}/(M_{\rm g}+M_{\rm *}+M_{\rm BH})$, as a function of redshift for our two simulations.
As seen in the cosmic SFR, at all times, the simulation without AGN has a higher stellar fraction than when AGN feedback is included.
AGN feedback also delays the redshift by which half of the present-day stellar mass has formed, from $z\sim1.5$ to $z\sim1.3$, a difference of about $0.5$ Gyr.
The $z=0$ values of $f_{\rm *}=0.079$ and $f_{\rm *}=0.068$ are both larger than the observational estimate of $f_{\rm *}=0.046\pm0.01$ \citep{fukugita04}, which may be because our simulations show a clustering of galaxies.
Our simulated data follow the same trend as observations, but lie slightly above observations at $z\sim2-3$.
Observational data are taken from the compilation of cosmic stellar densities in \citet{madau14}, converted to a mass fraction assuming $\Omega_{\rm b}=0.046$.

AGN feedback also greatly affects the chemical enrichment of the Universe.
In Fig. \ref{fig:4paper4_fg_all}c, we show how the gas oxygen abundance changes over the course of the simulations, separately considering all gas, the ISM, and the IGM.
Following \citet{ck07}, we define the ISM as being any gas particle belonging to a galaxy identified by a Friend of Friends group finder \citep{springel01}, and the IGM as all other gas particles.
In all cases, the simulation with AGN shows lower [O/H]$_{\rm g}$ at high redshift than the other simulation, and the opposite towards lower redshift.
At all redshifts, star formation is suppressed by AGN feedback, and the lower [O/H]$_{\rm g}$ at $z\gtsim3$ is due to the reduced enrichment of gas from stars.
However, at $z\ltsim2$, [O/H]$_{\rm g}$ is higher with AGN than without it, which is due to the metal ejection by AGN-driven winds.
This difference is more clearly seen in the IGM; by this time, BHs have grown massive enough to drive gas out of their galaxy and into the IGM, causing it to become more enriched than the case without AGN feedback \citepalias[for more details, see][]{pt15b}.
We also compare to observations; the data of \citet{balestra07} and \citet{rafelski12} (green triangles and blue squares, respectively) trace the intra cluster medium, which should lie between our simulated values for the ISM and IGM and follow a similar trend with redshift.
Data from \citet{aguirre08,ryanweber09,simcoe11,becker11} are for the metallicity of the IGM, and are in good agreement with our simulation with AGN feedback.

The influence of AGN feedback on the evolution of stellar metallicity is much more straightforward; Fig. \ref{fig:4paper4_fg_all}d shows this as a function of redshift, with the simulation without AGN having higher average metallicity at all times.
This difference is most pronounced at high redshift since AGN feedback delays early star formation and enrichment, and shifts the redshift at which half the present day metallicity is attained from $z=3.76$ to $z=3.07$, a difference of $\sim440$ Myr.
Towards low redshift the difference in metallicities is smaller, but since fewer stars are produced overall when AGN feedback is included, the enrichment is not as much by the present day.
The observational estimate of \citet{gallazzi08} at the present day lies slightly below the results of both simulations, which, again, is due to our initial conditions.

\section{Mass and Luminosity Functions}
\label{sec:4paper4_mfn}

\begin{figure*}
\centering
\includegraphics[width=\textwidth,keepaspectratio]{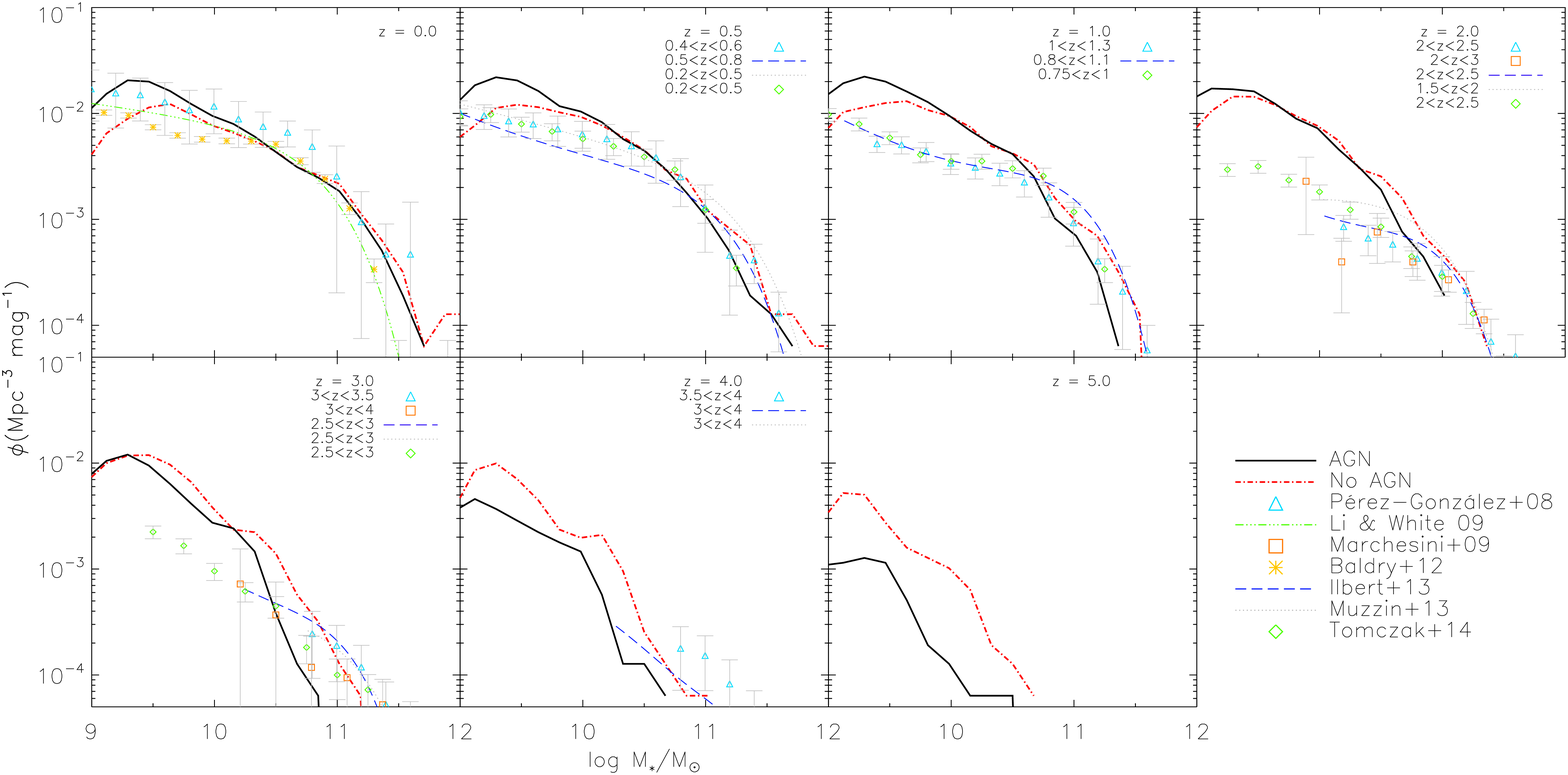}
\caption{Evolution of the galaxy stellar mass function.
Simulated data are shown as the solid black and red dot-dashed lines (with and without AGN respectively).
The observational data from \citet{perez08}, \citet{marchesini09}, \citet{baldry12}, and \citet{tomczak14} are shown, as well as the Schechter function fits of \citet{li09}, \citet{ilbert13}, and \citet{muzzin13}.}
\label{fig:4paper4_mf}
\end{figure*}

The stellar mass/luminosity functions are also important constraints for galaxy simulations, but there is limitation of the simulations volume, which should be taken into account when comparing to observations.
Fig. \ref{fig:4paper4_mf} shows the evolution of the stellar mass function of simulated galaxies at $z=0-5$, which has a similar shape at all redshifts.
At all redshifts, there are fewer high-mass galaxies when AGN feedback is included (solid lines), which is easily understood due to the suppression of star formation (e.g., \citetalias{pt15a}).
At the low mass end, the effect of AGN is not so simple.
There are more low-mass galaxies with AGN feedback at all but the highest redshift.
AGN feedback delays early star formation in all galaxies by heating the ISM.
This reduces the strength of supernova feedback, which would otherwise disrupt some low mass galaxies.
More low mass galaxies therefore survive to low redshift.
In addition, metal ejection to the IGM by AGN feedback could trigger star formation in low mass halos near massive galaxies.
We should note that the excess of low mass galaxies could be due to the limited simulation volume; our simulation box contains a cluster of galaxies, producing more sites for galaxy formation than might be expected for different initial conditions.

As was discussed in detail in \citetalias{pt15a}, at the present, the simulated mass function agrees fairly well with observational data of \citet{perez08}, \citet{li09}, and \citet{baldry12} except the under-abundance around the observed value of $M^*$.
This is also true at $z=0.5$, where we compare with the observational data of \citet{perez08}, \citet{ilbert13}, \citet{muzzin13}, and \citet{tomczak14}.
There is better agreement with the observed high-mass end at this redshift than $z=0$, but the low-mass slope is still too steep.
At $z=1$, both simulations, particularly the one that includes AGN feedback, under-predict the number of $M^*$ galaxies and overpredict the abundance of low-mass galaxies ($M_*\ltsim10^{10}\msun$).
The discrepancy at low masses is even more pronounced at $z=2$ compared with the observations of \citet{marchesini09} and \citet{tomczak14}; the observed normalization, $\phi^*$, decreases by $\sim0.5$ dex between $z=1$ and 2, while the simulated value remains unchanged.
The high-mass end, on the other hand, is well-reproduced, especially by the simulation without AGN feedback.
This is true also at $z=3$, although there is greater scatter between different observational datasets, and even though the simulated $\phi^*$ has decreased, there are still more low-mass galaxies than observed.
At high redshift, $z=4$, observations can constrain only the high-mass end of the mass function, and the simulated data lie below the observations of \citet{perez08} and \citet{ilbert13}.

From $z=4-2$, there is some evolution in $\phi^*$, and there is strong evolution in $M^*$ at all epochs in both simulations.
These trends are opposite to the observed strong evolution in $\phi^*$ and relative constancy of $M^*$ \citep[e.g.,][]{perez08,ilbert13,tomczak14}, but similar to the semi-analytic simulations of \citet{guo11} and \citet{guo13lf} who also find that low-mass galaxies form too early compared to observations.
The over-production of low mass galaxies has also been seen in other simulations \citep[e.g.,][]{oppenheimer06,dave11,hirschmann14}, where it was attributed to the inadequacy of the model for the production of stellar winds.
It might therefore be the case that our prescription for stellar feedback needs altering, or that BHs are exerting too much influence on the early stages of these galaxies due to our relatively simplistic model.
It is also possible, however, that UV background radiation suppresses the formation of low mass galaxies, which needs properly saving using radiative transfer.

\begin{figure*}
\centering
\includegraphics[width=\textwidth,keepaspectratio]{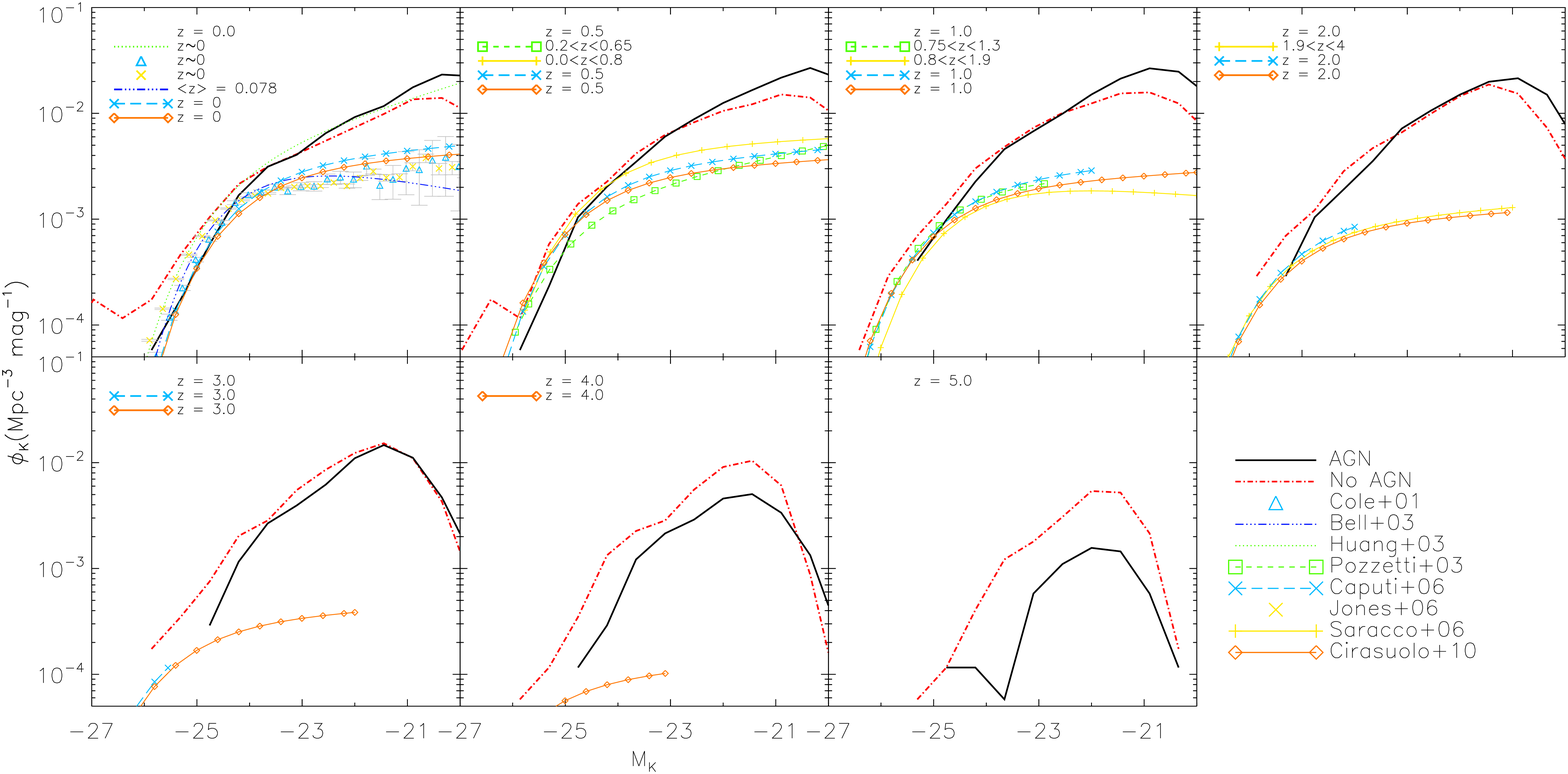}
\caption{Evolution of the rest-frame $K$-band luminosity function.
Simulated data are shown as the solid black and red dot-dashed lines (with and without AGN respectively).
Observational data from \citet{cole01} and \citet{jones06} are shown, as well as the Schechter function fits of \citet{bell03}, \citet{huang03}, \citet{pozzetti03}, and \citet{saracco06}, and the evolving Schechter function fits (see text) of \citet{caputi06} and \citet{cirasuolo10}.}
\label{fig:4paper4_lfk}
\end{figure*}

Related to the mass function, we show in Fig. \ref{fig:4paper4_lfk} the evolution of the $K$-band galaxy luminosity function of our simulations at $z=0-5$.
Note that, different from \citetalias{pt15a}, we calculate the luminosity of each galaxy simply by adding the contribution of each star particle in its FoF group, and not by estimating the total luminosity by fitting a S\'ersic function.
This means that all galaxies identified in the simulation contribute to the luminosity function, and since galaxy light is centrally concentrated the magnitudes obtained are fairly consistent by the two different methods.
This has the effect that, at ever lower masses, more galaxies contribute to the luminosity function than in \citetalias{pt15a}, but the differences between simulations found in that paper remain, and none of our conclusions is compromised.

At $z=0$, the high-mass end of the simulated data with AGN shows very good agreement with the observations of \citet{cole01} and \citet{jones06}, and the Schechter function fits to the data of \citet{bell03} and \citet{huang03}, while the simulation without AGN greatly over-predicts the number of such galaxies.
\citet{caputi06} and \citet{cirasuolo10} fit an evolving Schechter function to their data at various redshifts with
\begin{equation}\label{eq:4paper4_schec1}
M_K^*\left(z\right)=M_K^*\left(0\right)-\left(\frac{z}{z_M}\right)^{k_M},
\end{equation}
\begin{equation}\label{eq:4paper4_schec2}
\phi^*\left(z\right)=\phi^*\left(0\right)\exp\left[-\left(\frac{z}{z_\phi}\right)^{k_\phi}\right],
\end{equation}
and fixed low-mass slope $\alpha^*$.
The fitting parameters are used in each of the panels in Fig. \ref{fig:4paper4_lfk}, and the luminosity function plotted in the approximate range of magnitudes for which data were available.
There is also good agreement between the simulation with AGN at the high-mass end and these fits at $z=0$.
Both our simulations over-predict the number of low-mass galaxies compared to observations, except those of \citet{huang03}, who find a much steeper low-mass slope ($\alpha^*=-1.37$) than contemporaneous and subsequent studies ($\alpha^*\sim-1$).

By $z=0.5$, neither the simulated nor the observed luminosity functions have changed much compared to $z=0$, and the simulation without AGN still vastly over-predicts the numbers of the most luminous galaxies.
At $z=1$, the observed decline of $\phi^*$ with redshift is apparent, while no such change is seen in the simulations.
This trend continues out to $z=5$, with both simulations over-predicting the number of galaxies at all luminosities compared to observations.
The simulation with AGN produces consistently fewer galaxies at a given magnitude than the other simulation, at all but the lowest luminosities.

The constancy of the normalization, $\phi^*$, across a wide range of redshifts is clear, as is the increase of $M_K^*$ to lower luminosities at higher redshifts.
These trends lie in opposition to observations (but are similar to the evolution in the mass function); the fitted values of $k_M$ and $k_\phi$ of \citet{caputi06} and \citet{cirasuolo10} imply only weak evolution of $M_K^*$ with redshift, but relatively strong evolution of $\phi^*$.
 
The fact that there is not much evolution in the luminosity function in our simulations from high redshift implies that galaxies grow hierarchically and do not show downsizing.
However, we know from figure 10 of \citetalias{pt15a} that the stars that form the most massive galaxies at $z=0$ are very old in the simulation with AGN, while low-mass galaxies are, on average, much younger.
The large number of low-mass galaxies at high redshift may therefore be the progeny of today's most massive galaxies, with more low-mass galaxies forming at low redshift, leading to the constant normalization of the luminosity function.

\begin{figure*}
\centering
\includegraphics[width=\textwidth,keepaspectratio]{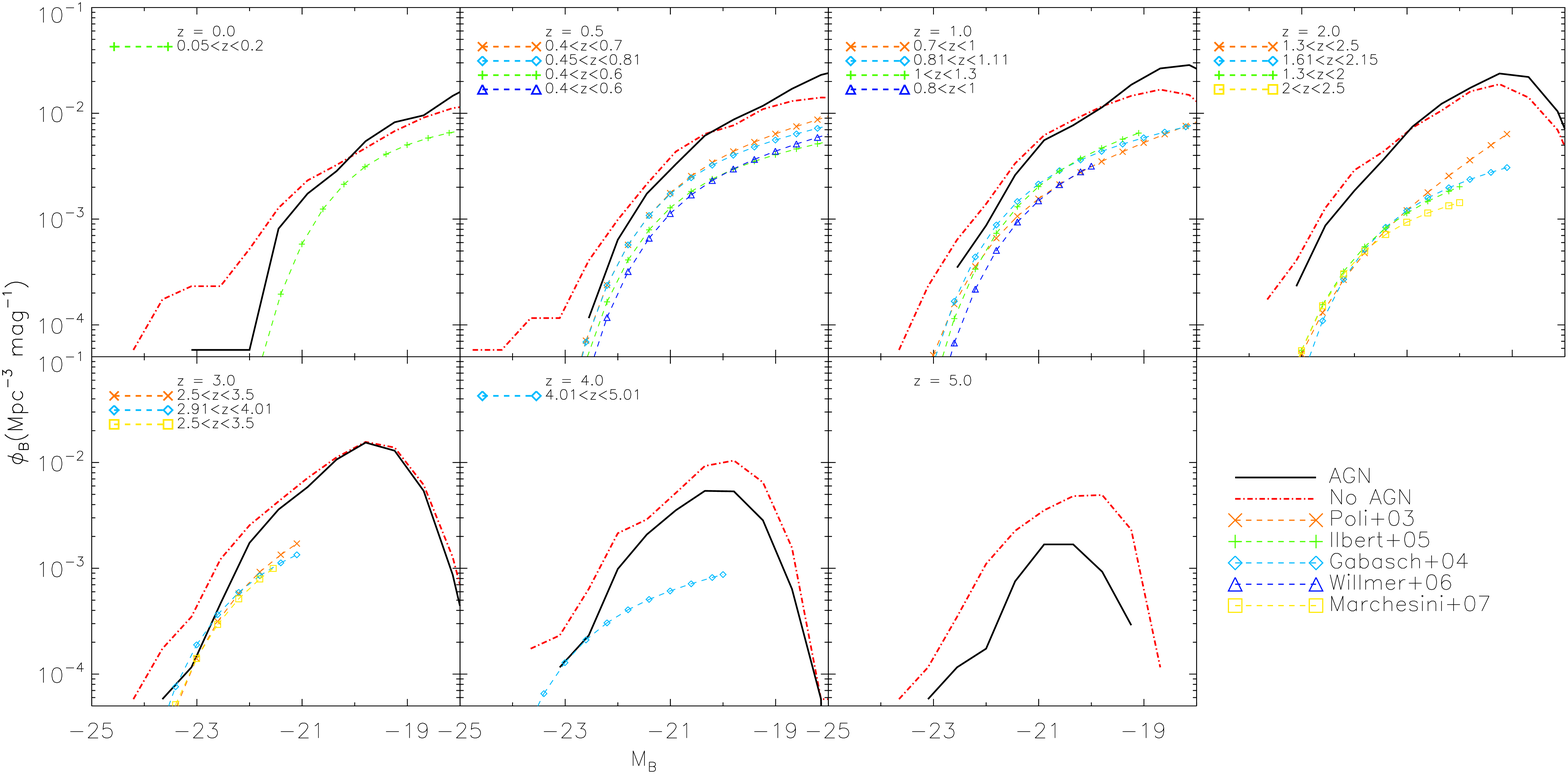}
\caption{Evolution of the rest-frame $B$-band luminosity function.
Simulated data are shown as the solid black and red dot-dashed lines (with and without AGN respectively).
The Schechter function fits to the observational data of \citet{poli03}, \citet{gabasch04}, \citet{ilbert05}, \citet{willmer06}, and \citet{marchesini07} are also shown.}
\label{fig:4paper4_lfb}
\end{figure*}

Fig. \ref{fig:4paper4_lfb} shows the evolution of the simulated $B$-band luminosity function at $z=0-5$.
It is worth noting that the $B$ band is more affected by dust extinction in observations than is $K$, but we do not include a prescription for dust extinction in our simulations.
All $M_B^*$ from observations have had 0.09 added to them to convert from the AB to Vega magnitude system \citep{blanton07}.

As with the $K$-band luminosity function, the simulation with AGN feedback predicts fewer high-luminosity galaxies than the simulation without at all redshifts, and both produce more low luminosity galaxies than are observed at all redshifts.
At redshifts away from the peak of star formation at $z=2$, the simulation with AGN feedback shows good agreement at the high luminosity end, but over-predicts the number of high luminosity galaxies at $z=1-2$.
This coincides with the epoch of maximum dust attenuation \citep{cucciati12}, which may affect the observed luminosity functions.

Applying a Schechter function to the simulated data shows that $M^*_B$ tends to decrease with increasing redshift, in qualitative agreement with observations.
The $B$ band traces recent star formation, which is quenched in the most massive galaxies at low redshift, leading to the observed evolution.
The decrease of $\phi_B^*$ is slower than observed, and only apparent from $z\sim2$.

\section{Line Fitting for $M_{\rm BH}$--$\sigma$}
\label{app:lineaverage}

As described in the main text, we fit straight lines to our simulated $M_{\rm BH}$--$\sigma$ relation assuming separately that $M_{\rm BH}$ and $\sigma$ is the independent variable.
Thus we obtain two fitted lines:
\begin{equation}\label{eq:la1}
	\log M_{\rm BH} = \alpha_1 + \beta_1 \log(\sigma/200\,{\rm kms}^{-1}),
\end{equation}
and
\begin{equation}\label{eq:prime}
	\log(\sigma/200\,{\rm kms}^{-1}) = \alpha' + \beta'\log M_{\rm BH}.
\end{equation}
Equation \eqref{eq:prime} may be rearranged to give
\begin{equation}\label{eq:la2}
	\log M_{\rm BH} = \alpha_2 + \beta_2 \log(\sigma/200\,{\rm kms}^{-1}),
\end{equation}
with $\alpha_2 = -\alpha'/\beta'$ and $\beta_2 = 1/\beta'$.
We now wish to find $\alpha$ and $\beta$ such that the line $\log M_{\rm BH} = \alpha + \beta \log(\sigma/200\,{\rm kms}^{-1})$ passes through the intersection of the lines defined by equations \eqref{eq:la1} and \eqref{eq:la2}, and makes an equal angle to both.
The second requirement suggests that
\begin{equation}
	\tan^{-1}\beta = \frac{1}{2}(\tan^{-1}\beta_1 + \tan^{-1}\beta_2),
\end{equation}
and hence that
\begin{equation}\label{eq:la5}
	\begin{split}
		\beta & = \frac{\sin\tan^{-1}\beta_1 + \sin\tan^{-1}\beta_2}{\cos\tan^{-1}\beta_1 + \cos\tan^{-1}\beta_2} \\
			& = \frac{\beta_1\sqrt{1+\beta_2^2} + \beta_2\sqrt{1+\beta_1^2}}{\sqrt{1+\beta_1^2}+\sqrt{1+\beta_2^2}}\\
			& \equiv \beta_1w_2 + \beta_2w_1,
	\end{split}
\end{equation}
where we have defined $w_i = \sqrt{1+\beta_i^2}/(\sqrt{1+\beta_1^2}+\sqrt{1+\beta_2^2})$.

We can now find expressions for $\log M_{\rm BH}$ and $\log(\sigma/200\,{\rm kms}^{-1})$ in terms of $\alpha_1$, $\alpha_2$, $\beta_1$, and $\beta_2$ from equations \eqref{eq:la1} and \eqref{eq:la2}, and substitute into our desired relation:
\begin{equation}
	\frac{\alpha_1\beta_2 - \alpha_2\beta_1}{\beta_1-\beta_2} = \alpha + \beta\frac{\alpha_2-\alpha_1}{\beta_1-\beta_2}.
\end{equation}
Substituting the expression for $\beta$ from equation \eqref{eq:la5}, and after some algebra, we find that
\begin{equation}
	\alpha = \alpha_1w_2 + \alpha_2w_1.
\end{equation}

\bsp

\label{lastpage}

\end{document}